\documentclass[%
reprint,	% True two column version
superscriptaddress,
showpacs,preprintnumbers,
amsmath,amssymb,
floatfix,
 prb,
]{revtex4-1}
\usepackage{graphicx}% Include figure files
\usepackage{dcolumn}% Align table columns on decimal point
\usepackage{bm}% bold math
\begin{document}

%\preprint{APS/123-QED}

\title{First-principles method for electron-phonon coupling and electron mobility: Applications to 2D materials}
%{Mobility and bulk electron-phonon interaction in two-dimensional materials}

\author{Tue Gunst}
\email{Tue.Gunst@nanotech.dtu.dk}
\affiliation{Department of Micro- and Nanotechnology (DTU Nanotech), Center for Nanostructured Graphene (CNG), Technical University of Denmark, DK-2800 Kgs. Lyngby, Denmark}\author{Troels Markussen}
\affiliation{QuantumWise A/S, Fruebjergvej 3, Postbox 4, DK-2100 Copenhagen, Denmark}
\author{Kurt Stokbro}
\affiliation{QuantumWise A/S, Fruebjergvej 3, Postbox 4, DK-2100 Copenhagen, Denmark}
\author{Mads Brandbyge}
\affiliation{Department of Micro- and Nanotechnology (DTU Nanotech), Center for Nanostructured Graphene (CNG), Technical University of Denmark, DK-2800 Kgs. Lyngby, Denmark}

\date{\today}
%Use showkeys class option if keyword display desired

\begin{abstract}
%\boldmath

We present density functional theory calculations of the phonon-limited mobility in n-type monolayer graphene, silicene and MoS$_2$.
The material properties, including the electron-phonon interaction, are calculated from first-principles.
We provide a detailed description of the normalized full-band relaxation time approximation for the linearized Boltzmann transport equation (BTE) that includes inelastic scattering processes.
The bulk electron-phonon coupling is evaluated by a supercell method. The method employed is fully numerical and does therefore not require a semi-analytic treatment of part of the problem and, importantly, it keeps the anisotropy information stored in the coupling as well as the band structure. 
In addition, we perform calculations of the low-field mobility and its dependence on carrier density and temperature to obtain a better understanding of transport in graphene, silicene and monolayer MoS$_2$.
Unlike graphene, the carriers in silicene show strong interaction with the out-of-plane modes. We find that graphene has more than an order of magnitude higher mobility compared to  silicene. For MoS$_2$, we obtain several orders of magnitude lower mobilities in agreement with other recent theoretical results. The simulations illustrate the predictive capabilities of the newly implemented BTE solver applied in simulation tools based on first-principles and localized basis sets.
%in the Atomistix ToolKit (ATK) simulation tool.
%first-principles theoretical methods
%In the case of silicene, the results illustrate that the out-of-plane acoustic phonon mode may play the dominant role unlike its close relative, graphene.
%We find good agreement with experiments... examples.
%For monolayer MoS2 we obtain. This value is higher than what is observed experimentally but is on the same order of magnitude as other recent theoretical results.

%We provide a detailed description of the full bandstructure method employed
%full bandstructure method employed and describe a
\end{abstract}

\maketitle
%\tableofcontents

\section{Introduction}
Two-dimensional (2D) materials are promising candidates for future electronic devices.\cite{akinwande_two-dimensional_2014,geim_van_2013,efetov_controlling_2010,tao_silicene_2015,xu_graphene-like_2013,butler_progress_2013}
Examples of three such materials are illustrated in Fig.1, showing a monolayer of graphene (A), its Silicon based counterpart, silicene\cite{feng_evidence_2012,kara_review_2012,houssa_silicene:_2015,balendhran_elemental_2015} (B), and monolayer MoS$_2$ (C). In such systems two-dimensionality allows very precise control of the carrier density by a gate which enables tuning of the electron-phonon interaction.\cite{efetov_controlling_2010} Electron-phonon interaction in graphene has been studied previously\cite{hwang_acoustic_2008,chen_diffusive_2009,kaasbjerg_unraveling_2012}, but with the recent advances in fabrication of devices based on other 2D materials, like the first demonstration of a silicene transistor\cite{tao_silicene_2015}, further studies of interaction phenomena in 2D materials are necessary.
\begin{figure}[!htbp]%[!t]%[!htbp]%
\centering
{\includegraphics[width=0.72\linewidth]{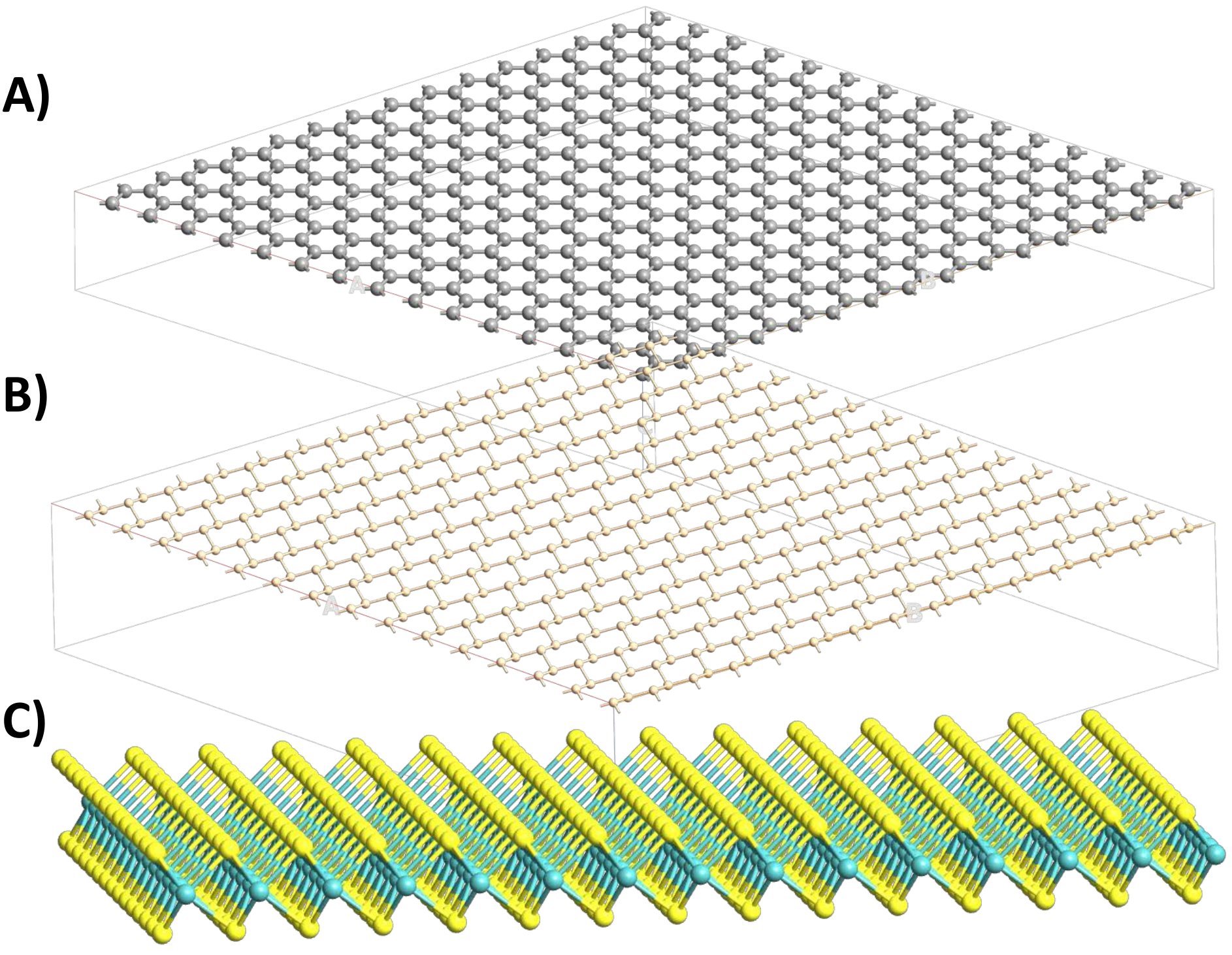}}%width=0.99\linewidth%width=2.5in
\caption{The monolayered two-dimensional systems that are considered are A) graphene, B) silicene and C) MoS$_2$.}
\label{fig:Systems}
\end{figure}

%The efficiency of electrons in a given material to respond to a applied electric field is often measured by the mobility.
%Mobility is important 
When comparing the electrical performance of devices one often considers the carrier mobility of the materials.
Mobility is a key parameter for the semiconductor industry describing the motion of electrons when an electric field is applied.
Experiments can approach the 'intrinsic' phonon-limited mobility by several means. One experiment combine defect-free edge contacting\cite{wang_one-dimensional_2013} of gate-tunable graphene electrodes with MoS$_2$ encapsulated in hexagonal boron nitride layers.\cite{cui_multi-terminal_2015} Several other experiments screen the scattering from charged impurities by a high-$\kappa$ gate dielectrics\cite{radisavljevic_single-layer_2011} or through device suspension in high-$\kappa$ liquids.\cite{newaz_probing_2012,shishir_intrinsic_2009,chen_ionic_2009}
In general, van der Waals heterostructures may pave the way for devices with reduced extrinsic scattering, like charged impurity scattering, rendering modeling of electron-phonon scattering in these devices even more important.
%Experiments are approaching the intrinsic mobility limits for instance by combining defect free edge contacting of gate-tunable graphene electrodes with MoS$_2$ encapsulated in hexagonal boron nitride layers\cite{cui_multi-terminal_2015} or by screening of charged impurities by high-$\kappa$ gate dielectrics\cite{radisavljevic_single-layer_2011} or device suspension in high-$\kappa$ liquids.\cite{newaz_probing_2012,shishir_intrinsic_2009}

Conventional mobility modeling usually considers effective mass approximations\cite{ando_electronic_1982} in the case of semiconductors or linear bands for semimetals\cite{stauber_electronic_2007,hwang_acoustic_2008,manes_symmetry-based_2007} combined with empirical deformation potentials and semi-analytical solutions of the Boltzmann equation.\cite{hwang_acoustic_2008,das_sarma_electronic_2011} The Boltzmann theory is a useful approach to model the low-field/linear response mobility\cite{kawamura_phonon-scattering-limited_1992,fischetti_band_1996,das_sarma_electronic_2011}, as well as the high-field transport through Monte Carlo simulations.\cite{fang_high-field_2011,shishir_velocity_2009,fischetti_monte_1993}
Several studies have examined effects related to screening\cite{hwang_screening-induced_2009,kuroda_nonlinear_2011}, scattering from out-of-plane vibrations\cite{mariani_flexural_2008} and performed atomistic calculations of the mobility from tight-binding\cite{sule_phonon-limited_2012,niquet_fully_2012,cauley_distributed_2011} as well as electron-phonon interactions role in facilitating interlayer conduction\cite{perebeinos_phonon-mediated_2012} and current-induced heating.\cite{gunst_phonon_2013,lu_current-induced_2012}
Density functional theory (DFT) and atomistic methods can be used to assess the electronic structure and electron-phonon coupling in novel 2D materials where fitted deformation potential parameters are not available.\cite{park_electronphonon_2014,sohier_phonon-limited_2014,wang_thermoelectric_2011,restrepo_first-principles_2009,li_intrinsic_2013,shao_first-principles_2013,szabo_textitab_2015}

Recently, several groups have combined DFT with density functional perturbation theory\cite{borysenko_first-principles_2010,restrepo_first_2014,park_electronphonon_2014,sohier_phonon-limited_2014,yan_electron-phonon_2013} to evaluate the mobility from first-principles. Parameter-free methods can be used to address how close experiments are to ideal conductivities and if further optimization of fabrication techniques and device designs for novel 2D materials could improve device performance.
In addition, first-principles calculations of the bulk electron-phonon interaction may be used for comparing deformation potential values to those obtained from experiments\cite{yoder_abinitio_1993,kaasbjerg_acoustic_2013} and used to conclude which scattering effects are dominant.
The dominant effect is not always directly clear from experiments and published deformation potentials can vary significantly.\cite{hwang_acoustic_2008}

Several techniques which differ from contacted electrical measurements are in development that can provide detailed knowledge of the electron-phonon interaction; broadening of Raman peaks\cite{lazzeri_phonon_2006}, kinks in the angle-resolved photoemission spectra\cite{mazzola_kinks_2013} and non-destructive optical methods.\cite{buron_graphene_2015} Hereby a first-principles method to evaluate bulk electron-phonon interactions may provide useful data for comparison with experiments.

%Bulk eph:
Methods to obtain the bulk electron-phonon interactions include Wannier functions together with a generalized Fourier interpolation scheme\cite{giustino_electron-phonon_2007}, perturbation theory and empirical pseudopotentials\cite{fischetti_pseudopotential-based_2013}, finite-differences in the projector-augmented wave method\cite{kaasbjerg_phonon-limited_2012} and density functional perturbation theory.\cite{piscanec_kohn_2004,borysenko_first-principles_2010,restrepo_first_2014,li_electrical_2015}
%supercell approach analogous to that used for the calculation of the phonon dispersion%that can be used in a 

We will later give a simple derivation of an expression for the bulk electron-phonon coupling applicable in any setup based on localized basis sets.
From this we have implemented a supercell method used to calculate the bulk electron-phonon interaction employing finite differences. This method is analogous to that used for calculating the phonon dispersion both in terms of methodology and computational cost.
%We shall later give a simple derivation of the supercell method used to calculate the bulk electron-phonon interaction

In this paper, we present a detailed description of the implementation of the bulk electron-phonon coupling, and a Boltzmann Transport Equation (BTE) solver in the Atomistix ToolKit (ATK) simulation tool.\cite{ATK} We apply atomistic simulations with ATK to study electron-phonon coupling in 2D materials from first-principles.
We formulate a normalized full-band relaxation time approximation (RTA) for the linearized BTE that includes inelastic scattering processes.
The bulk electron-phonon methodology employed makes simulations possible that do not require a semi-analytic treatment of part of the problem and, importantly, we keep the anisotropy information stored in the coupling as well as the band structure. 
In addition, we perform calculations of the low-field mobility and its dependence on carrier density and temperature to obtain a better understanding of transport in graphene, silicene and MoS$_2$.% and other 2D materials.

Despite the fact that several papers have presented schemes  for obtaining electron-phonon coupling and mobilities from first-principles calculations, only few applications exist and the methods are by no means standard. 
The focus of our work is to make an efficient and practical scheme for such calculations in order to make them as accessible as performing a bandstructure calculation. In the methods section we will discuss the differences in our technical implementation compared to previous work, and  here we only briefly mention some general advantages of our work which we believe will be important for the wide spread usage.  
The method is based on localized basis sets which allows for exploiting locality in real space and we have made a great effort of optimizing the implementation and apply efficient parallelization and interpolation schemes. The method is therefore fast and able to handle large systems. Secondly, it is highly accurate and as we will show in this paper, it gives results consistent with previous DFT simulations. Finally, it is implemented in a versatile framework with an easy to use python interface which allows for performing all the different parts of the calculations with a single script, thus, it is easy to setup the calculation and it requires minimal human interference for performing the calculation.
%Despite the fact that several papers discuss approaches for obtaining electron-phonon coupling and mobilities from first-principles calculations, only few applications exist and the methods are by no means standard.
%In addition to the methodically differences in the implementation discussed in the methods section, the main advantages with our implementation is outlined as follows.
%Firstly, it is based on localized basis sets and therefore is fast and able to handle large systems. Secondly, as we will show in this paper, it gives results consistent with previous DFT simulations. Finally, it is implemented in a versatile framework with an easy-to-use python interface which should 
%extend the field of applications and pave the way for a faster and more practical assessment of transport properties of emerging nanomaterials.
%In addition, we have made a great effort of optimizing the implementation and apply efficient parallelization and interpolation schemes.
%We therefore believe that this tool will simplify the simulation
%in a few key points.
%unified treatment of electrons and phonons in both bulk systems and larger nanostructures.
%pave the way for a much wider
%field of applications
% application of the methods in the future.

The paper is organized as follows.
In Sec.~\ref{Sec:Method} we present the theoretical and numerical methods
used. We derive expressions for the linearized BTE and the bulk electron-phonon interaction implemented in the ATK simulation tool.
In Sec.~\ref{Sec:Simulations} we present our results for the bulk electron-phonon coupling in graphene, silicene and MoS$_2$. In addition, we discuss the dependence of the mobility on the carrier density and temperature for all three materials.
Finally, the results are summarized and discussed in Sec.~\ref{Sec:Conclusion}.

%------ put before paragraph above...
% Review of graphene, silicene and MoS2 research.

% Cite DFT for deformation potential.
% Boltzmann theory
% We use the Boltzmann transport theory
% Beyond deformation potential - The bulk electron-phonon interaction is used directly in the transport calculations. However, it may also be used to fitting deformation potential parameters which can be used in semi-analytic approximations or kinitic Monte Carlo device simulations.\cite{...} 

%graphene on SiO2: chen_diffusive_2009
%analytical approach towards electron-phonon coupling in graphene\cite{manes_symmetry-based_2007}
%mobility of silicene\cite{tao_silicene_2015}
%measurements of graphene mobility\cite{chen_diffusive_2009,efetov_controlling_2010}

%MoS2: experiments suggest a room temperature mobility of MoS$_2$ of $\approx 200\,\mathrm{cm}^2/\mathrm{V}\,\mathrm{s}$. \cite{radisavljevic_single-layer_2011,fivaz_mobility_1967,kaasbjerg_phonon-limited_2012}
% Graphene:
% Silicene: $\approx 100\,\mathrm{cm}^2/\mathrm{V}\,\mathrm{s}$.\cite{tao_silicene_2015}

\section{Methods\label{Sec:Method}}
In the diffusive transport limit, the mobility can be obtained by solving the semiclassical BTE for the electronic distribution function, $f(\varepsilon_{\mathbf{k}n}) = f_{\mathbf{k}n}$:
\begin{eqnarray}
\frac{\partial f_{\mathbf{k}n}}{\partial t}
+ \mathbf{v}_{\mathbf{k} n} \cdot \nabla_{\mathbf{r}} f_{\mathbf{k} n}
+\frac{\mathbf{F}}{\hbar} \cdot \nabla_{\mathbf{k}} f_{\mathbf{k} n}
= \left.\frac{\partial f_{\mathbf{k}n}}{\partial t}\right|_{coll}\,.\label{FullBTE}
\end{eqnarray}
Here $\mathbf{k}, n$ labels the k-point and band index, respectively. The velocity is defined as $\mathbf{v}_{\mathbf{k} n}=1/\hbar \nabla_{\mathbf{k}} \varepsilon_{\mathbf{k}n}$ and $\mathbf{F}=q (\mathbf{E}+\mathbf{v}\times \mathbf{B})$ gives the external force. The right hand side in Eq.~\eqref{FullBTE} describes different sources of scattering and dissipation that drives the system towards steady state.
In the case of a homogeneous system, zero magnetic field, and a time-independent electric field in the steady state limit, the BTE simplifies to:
\begin{eqnarray}
\frac{q\mathbf{E}}{\hbar} \cdot \nabla_{\mathbf{k}} f_{\mathbf{k} n}
= \left.\frac{\partial f_{\mathbf{k}n}}{\partial t}\right|_{coll} \,.\label{SteadyStateBTE}
\end{eqnarray}
Assuming instantaneous, single collisions, which are independent of the driving force, the collision integral can be expressed using transition rates, $P_{\mathbf{k}\mathbf{k}'}^{nn'}$,
\begin{eqnarray}
\left.\frac{\partial f_{\mathbf{k}n}}{\partial t}\right|_{coll} = &-& \sum_{\mathbf{k}'n'} \left[f_{\mathbf{k}n}\left(1-f_{\mathbf{k}'n'}\right)P_{\mathbf{k}\mathbf{k}'}^{nn'}\right.
\nonumber\\
&-&\left.f_{\mathbf{k}'n'}\left(1-f_{\mathbf{k}n}\right)P_{\mathbf{k}'\mathbf{k}}^{n'n}\right] \,. \label{eqn:collIntegral}
\end{eqnarray}
The transition rate due to phonon scattering from a state $|\mathbf{k}n\rangle$ to $|\mathbf{k}'n'\rangle$ is obtained from Fermi's golden rule (FGR):
\begin{eqnarray}
P_{\mathbf{k}\mathbf{k'}}^{nn'}&=&\frac{2\pi}{\hbar} \sum_{\mathbf{q},\lambda}|g_{\mathbf{k}\mathbf{q}}^{\lambda n n'}|^2 
\left[ n_{\mathbf{q}}^{\lambda} \delta \left(\epsilon_{\mathbf{k}'n'}-\epsilon_{\mathbf{k}n}-\hbar \omega_{\mathbf{q} \lambda} \right) \delta_{\mathbf{k}',\mathbf{k}+\mathbf{q}}\right.
\nonumber\\
&+&\left. (n_{\mathbf{-q}}^{\lambda}+1) \delta \left( \epsilon_{\mathbf{k}'n'}-\epsilon_{\mathbf{k} n}+\hbar \omega_{-\mathbf{q} \lambda} \right)  \delta_{\mathbf{k}',\mathbf{k}-\mathbf{q}} \right] \,,\label{eqn:FGRtransitionRate}
\end{eqnarray}
describing absorption (first term) and emission (last term) of a phonon. The last term includes spontaneous emission which remains at zero temperature. The sum runs over phonon momentum ($\mathbf{q}$) and phonon branch index ($\lambda$). In Eq.~\eqref{eqn:FGRtransitionRate} we have explicitly stated the momentum conservation coming from the bulk electron-phonon interaction matrix element. We have implemented a supercell method to calculate the bulk electron-phonon interaction, $g_{\mathbf{k}\mathbf{q}}^{\lambda n n'}$, employing finite differences in a localized basis set (derived in section.~\ref{Sec:BulkEph}).

We will assume unperturbed phonons and apply the equilibrium Bose-Einstein distributions, $n_{\mathbf{q}}^{\lambda}=n_{\mathbf{q}}^{0,\lambda}$, in which case $n_{\mathbf{-q}}^{0,\lambda}=n_{\mathbf{q}}^{0,\lambda}$ since $\omega_{-\mathbf{q} \lambda}=\omega_{\mathbf{q} \lambda}$.
The transition rates $P_{\mathbf{k}\mathbf{k'}}^{\lambda nn'}$ and $P_{\mathbf{k}'\mathbf{k}}^{\lambda n'n}$ are linked through the "detailed balance equation"\cite{kawamura_phonon-scattering-limited_1992},
\begin{eqnarray}
\left[f^0_{\mathbf{k}n}\left(1-f^0_{\mathbf{k}'n'}\right)P_{\mathbf{k}\mathbf{k'}}^{nn'}-f^0_{\mathbf{k}'n'}\left(1-f^0_{\mathbf{k}n}\right)P_{\mathbf{k}'\mathbf{k}}^{n'n}\right] = 0 \label{BTE_detailed_balance} \,.
\end{eqnarray}
Here $f^0$ is the equilibrium Fermi distribution function.
This equation secures that $\left.\frac{\partial f_{\mathbf{k}n}}{\partial t}\right|_{coll}=0$ in equilibrium.

We linearize the BTE in the electric field.
The left hand side of the BTE, Eq.~\eqref{SteadyStateBTE}, is approximated to linear order in the electric field by changing to the equilibrium distribution:
\begin{eqnarray}
\frac{q\mathbf{E}}{\hbar} \cdot \nabla_{\mathbf{k}} f_{\mathbf{k} n}
&\approx& \frac{q\mathbf{E}}{\hbar} \cdot \nabla_{\mathbf{k}} f^0_{\mathbf{k} n} = q \mathbf{E} \cdot \mathbf{v}_{\mathbf{k} n} \frac{\partial f^0_{\mathbf{k}n}}{\partial \epsilon_{\mathbf{k}n}} \,.\label{BTE_LHS}
\end{eqnarray}
The right hand side is linearized by assuming a form of the distribution function that is linear in the electric field.
Defining a generalized transport relaxation time\cite{sohier_phonon-limited_2014,kawamura_phonon-scattering-limited_1992}, $\tau_{\mathbf{k} n}$, so that 
\begin{eqnarray}
f_{\mathbf{k}n}=f^0_{\mathbf{k}n} + q\mathbf{E} \cdot \mathbf{v}_{\mathbf{k} n} \tau_{\mathbf{k} n} (-\frac{\partial f^0_{\mathbf{k}n}}{\partial \epsilon_{\mathbf{k}n}}) \,,\label{eqn:NonequiDistribution}
\end{eqnarray}
and combining and inserting Eqs.~\eqref{eqn:collIntegral}, \eqref{eqn:FGRtransitionRate}, \eqref{BTE_detailed_balance}, \eqref{BTE_LHS} in Eq.~\eqref{SteadyStateBTE}, we arrive at the linearized BTE,
\begin{eqnarray}
1 &=&
\sum_{\mathbf{k}'n'} P_{\mathbf{k}\mathbf{k'}}^{nn'} \frac{(1-f^0_{\mathbf{k}'n'})}{(1-f^0_{\mathbf{k}n})}\nonumber\\
&\times&\left[
\tau_{\mathbf{k}n} -
\tau_{\mathbf{k}'n'} \frac{n_{\mathbf{k}'n'}}{n_{\mathbf{k}n}} \frac{f^0_{\mathbf{k}n}(1-f^0_{\mathbf{k}n})}{f^0_{\mathbf{k}'n'}(1-f^0_{\mathbf{k}'n'})} \right] \,.\label{BTE_integral_eqn}
\end{eqnarray}
Where we defined the direction projections $n_{\mathbf{k}n} = \hat{\mathbf{E}} \cdot \hat{\mathbf{v}}_{\mathbf{k}n}$.
Equation~\eqref{BTE_integral_eqn} is still a full integral equation. However, several approximations, termed relaxation time approximations (RTA), exist throughout literature\cite{Mahan,park_electronphonon_2014,stanojevic_validity_2014} to reduce the problem to a $\mathbf{k}'$-space integration.
For instance the term in the brackets in Eq.~\eqref{BTE_integral_eqn} is replaced by $\tau_{\mathbf{k}n}$ times the non-normalized factors $\left[1-\frac{\mathbf{k}\cdot \mathbf{k}'}{k^2}\right]$, $\left[1-\frac{\mathbf{v}_{\mathbf{k}' n'} \cdot \mathbf{v}_{\mathbf{k} n}}{|\mathbf{v}_{\mathbf{k} n}|^2}\right]$
or the normalized factors $\left[1-\frac{\mathbf{k}\cdot \mathbf{k}'}{kk'}\right]$ and infrequently $\left[1-\frac{\mathbf{v}_{\mathbf{k}' n'} \cdot \mathbf{v}_{\mathbf{k} n}}{|\mathbf{v}_{\mathbf{k}' n'}| |\mathbf{v}_{\mathbf{k} n}|}\right]$.\cite{stanojevic_validity_2014} The non-normalized conditions are related to the assumption that $\tau_{\mathbf{k}'n'}\approx \tau_{\mathbf{k}n}$ while the normalized expressions are related to the assumption that $\tau_{\mathbf{k}'n'}  |\mathbf{v}_{\mathbf{k}' n'}|\approx \tau_{\mathbf{k}n}  |\mathbf{v}_{\mathbf{k} n}|$, and in both cases that the last Fermi-factor is equal to unity.
The expressions based on group velocities have the advantage that they do not depend on the chosen reference $\mathbf{k}$-point.
In addition, non-normalized expressions may lead to unphysical negative momentum relaxation times.
Here we define a normalized full-band RTA of the linearized BTE, including inelastic scattering processes, as:
%Defining a generalized transport relaxation time, $\tau_{\mathbf{k} n}$, so that $f_{\mathbf{k}n}=f^0_{\mathbf{k}n} + q\mathbf{E} \cdot \mathbf{v}_{\mathbf{k} n} \tau_{\mathbf{k} n} (-\frac{\partial f^0_{\mathbf{k}n}}{\partial \epsilon_{\mathbf{k}n}})$, 
%we rewrite the linearized BTE in the relaxation time approximation:
\begin{eqnarray}
\frac{1}{\tau_{\mathbf{k} n}} &=& \sum_{\mathbf{k}'n'} \frac{(1-f^0_{\mathbf{k}'n'})}{(1-f^0_{\mathbf{k}n})} \left(1-\rm{cos}(\theta_{\mathbf{k}\mathbf{k'}})\right) P_{\mathbf{k}\mathbf{k'}}^{nn'} \,.\label{eqn:RTA_rate}
\end{eqnarray}
Here the scattering angle is defined by\cite{Note1}
\begin{eqnarray}
\rm{cos}(\theta_{\mathbf{k}\mathbf{k'}}) =\frac{n_{\mathbf{k}'n'}}{n_{\mathbf{k}n}}= \frac{\mathbf{v}_{\mathbf{k}' n'} \cdot \mathbf{v}_{\mathbf{k} n}}{|\mathbf{v}_{\mathbf{k}' n'}| |\mathbf{v}_{\mathbf{k} n}|} \,.
\end{eqnarray}
For an angle-independent transition rate then small-angle scattering, where $\rm{cos}(\theta_{\mathbf{k}\mathbf{k'}})\approx 1$, does not obstruct the flow of electrons whereas large-angle scattering, where $\rm{cos}(\theta_{\mathbf{k}\mathbf{k'}})\approx -1$, significantly increases resistivity. However, the selection rules stored in the bulk electron-phonon coupling matrix element complicates this general trend.
The current density is related to the average velocity, $\mathbf{J}=q n_0 \langle \mathbf{v}_{\mathbf{k} n} \rangle$, obtained from the nonequilibrium distribution function in Eq.~\eqref{eqn:NonequiDistribution}. From the transport relaxation time, Eq.~\eqref{eqn:RTA_rate}, we then evaluate the low-field electron mobility\cite{Mahan}
\begin{eqnarray}
\mu_e = -2q \frac{\sum_{\mathbf{k}n\in c} |\mathbf{v}_{\mathbf{k} n}|^2 \frac{\partial f^0_{\mathbf{k}n}}{\partial \epsilon_{\mathbf{k}n}} \tau_{\mathbf{k} n}}{n_0} \,,%\nonumber \\
\label{BTE_mobility}
\end{eqnarray}
where a factor of two accounts for the spin degeneracy.
The hole mobility is obtained by replacing the electron carrier density, $n_0=\sum_{\mathbf{k}n\in c}f^0_{\mathbf{k}n}$, and the summation over conductions bands ($c$) with the hole carrier density, \\$p_0=\sum_{\mathbf{k}n\in v}(1-f^0_{\mathbf{k}n})$, and a summation over valance bands ($v$). The total conductivity is given by $\sigma=q n_0 \mu_e + q p_0 \mu_h$.

We also mention the difference of the transport scattering time discussed here versus the lifetime of electronic quasiparticles.
The lifetime can be measured by angle-resolved photoemission spectroscopy and is evaluated as
\begin{eqnarray}
\frac{1}{\tau^{l}_{\mathbf{k} n}} &=& \frac{2 \pi}{\hbar} \sum_{\mathbf{k}'\mathbf{q}n'\lambda} |g_{\mathbf{k}\mathbf{q}}^{\lambda n n'}|^2 \nonumber \\
&\times&
\left[
\left(n_{\mathbf{q}}^{\lambda}+f^0_{\mathbf{k}'n'}\right) \delta \left(\epsilon_{\mathbf{k}'n'}-\epsilon_{\mathbf{k}n}-\hbar \omega_{\mathbf{q} \lambda} \right)\delta_{\mathbf{k}',\mathbf{k}+\mathbf{q}} \right.\nonumber\\
&+&
\left.
\left(1+n_{\mathbf{q}}^{\lambda}-f^0_{\mathbf{k}'n'}\right) 
\delta \left( \epsilon_{\mathbf{k}'n'}-\epsilon_{\mathbf{k} n}+\hbar \omega_{\mathbf{q} \lambda} \right) \delta_{\mathbf{k}',\mathbf{k}-\mathbf{q}}
\right] \,.\nonumber 
\end{eqnarray}
Performing the $\mathbf{k}'$ sum we obtain
\begin{eqnarray}
\frac{1}{\tau^{l}_{\mathbf{k} n}} &=& \frac{2 \pi}{\hbar} \sum_{\mathbf{q}n'\lambda} |g_{\mathbf{k}\mathbf{q}}^{\lambda n n'}|^2 \left[
\left(n_{\mathbf{q}}^{\lambda}+f^0_{\mathbf{k}+\mathbf{q}n'}\right) \delta \left(\epsilon_{\mathbf{k}+\mathbf{q}n'}-\epsilon_{\mathbf{k}n}-\hbar \omega_{\mathbf{q} \lambda} \right) \right.\nonumber\\
&+&
\left.
\left(1+n_{\mathbf{q}}^{\lambda}-f^0_{\mathbf{k}-\mathbf{q}n'}\right) 
\delta \left( \epsilon_{\mathbf{k}-\mathbf{q}n'}-\epsilon_{\mathbf{k} n}+\hbar \omega_{\mathbf{q} \lambda} \right) \right] \,.
\label{eqn:lifetime}
\end{eqnarray}
This equation can be derived by keeping only all terms proportional to $f_{\mathbf{k}n}$ in Eq.~\eqref{eqn:collIntegral} and corresponds approximately to neglecting the scattering angle transport factor, $(1-\rm{cos}(\theta))$, in Eq.~\eqref{eqn:RTA_rate}.\cite{Mahan,hwang_single-particle_2008,kaasbjerg_unraveling_2012}

%$\sigma$ is the spin index
%...Inlcuding inelastic scattering processes.

\subsection{Numerical details}
The full $\mathbf{k}$-dependent scattering rate, in contrast to a simplified energy-dependent expression, leads to several numerical complications which has not been addressed in the literature to the best of our knowledge. Therefore, we will outline the main technicalities needed to account for the anisotropy information stored in the full band-structure and the electron-phonon coupling.
We represent the delta-functions in Eq.~\eqref{eqn:FGRtransitionRate} by Lorentzians, $\delta (\epsilon)\approx \frac{1}{\pi} \frac{\gamma/2}{(\gamma/2)^2 + \epsilon^2}$, with a finite broadening $\gamma$. Consequently, we find that it is important to evaluate the Fermi-factors $f^0_{\mathbf{k}'n'}$ at the exact final energy, $\epsilon_{\mathbf{k}n}\pm \hbar \omega$ as opposed to $\epsilon_{\mathbf{k\pm q}n}$, to get correct results. This also means that absorption and emission terms need to be handled independently when it comes to the peculiar Fermi-prefactor in Eq.~\eqref{eqn:RTA_rate}.
In addition, we rewrite the prefactor in the case of absorption, to avoid numerical instabilities:
\begin{eqnarray}
\frac{(1-f^0_{\mathbf{k}'n'})}{(1-f^0_{\mathbf{k}n})} = \frac{f^0_{\mathbf{k}n}}{f^0_{\mathbf{k}'n'}} \frac{1+n_B(\epsilon_{\mathbf{k}'n'}-\epsilon_{\mathbf{k}n})}{n_B(\epsilon_{\mathbf{k}'n'}-\epsilon_{\mathbf{k}n})} \label{eqn:fermi2boltzmann}\,.
\end{eqnarray}
This secures a stable denominator, since absorption will dominate the low energy spectrum, $\epsilon < \mu_F$  where $f^0 \rightarrow 1$ and $1-f^0 \rightarrow 0$.
Considering the $\mathbf{k}$- and $\mathbf{q}$-grids a conversion factor of $\Delta \mathbf{q}/\Delta \mathbf{k}$ is needed if the $\mathbf{k}$-grid and $\mathbf{q}$-grid are not equivalent. One can show that if the two grids are equivalent the full linearized BTE, Eq.~\eqref{BTE_integral_eqn}, simplifies to a linear matrix equation. However, it is advantageous to allow for different grid resolutions in order to apply smart-chosing of the grids and resulting simulation speedup. A fine resolution of the final state $\mathbf{q}$-grid secures a correct result for each $\mathbf{k}$-point even at a rough $\mathbf{k}$-grid. Our approach has therefore been to use fine $\mathbf{q}$-grids with the possibility of interpolation to even higher resolution for all $\mathbf{q}$-dependent variables.

% From $\mathbf{k}'$-grid to $\mathbf{q}$-grid. Conversion factor if $\mathbf{k}$-grid and $\mathbf{q}$-grids are not equivalent. One can show that if the two grids are equal the full BTE ...ref... simplifies to a linear matrix equation. However, it is advantageous to allow for different grid resolutions to allow for smart-chosing of the grids and resulting simulation speedup.
%Interpolation of q-grid: fine resolution of the final state q-grid secures a correct result for each k-point even at a rough k-grid. Our approach has therefore been to use fine q-grids that is also interpolated to even higher resolution.
%and in addition apply  of this grid for each calculation..Refinement...
%In addition, ... velocity. Perturbation theory vs. finite difference (sorting).\\
%The velocities, $\mathbf{v}_{\mathbf{k} n},\mathbf{v}_{\mathbf{k}' n'}$, are obtained from perturbation theory to avoid finite difference errors from crossing bands.
As a final remark we mention that the velocities, $\mathbf{v}_{\mathbf{k} n},\mathbf{v}_{\mathbf{k}' n'}$, are obtained from perturbation theory.
From a change $d \mathbf{k}_{\alpha}$, we obtain the derivative
\begin{eqnarray}
\frac{d \mathbf{H}_{\mathbf{k}}}{d \mathbf{k}_{\alpha}} = \sum_I i\mathbf{R}_{I,\alpha} \mathbf{H}_{\nu \mu} e^{i\mathbf{k}\cdot\mathbf{R}_I} \,,
\end{eqnarray}
of the Hamiltonian matrix in a basis of localized orbitals ($\nu,\mu$), and correspondingly for the overlap matrix $\mathbf{S}$ by replacing $\mathbf{H}\rightarrow \mathbf{S}$.
Here $I$ labels the unit cells, with lattice vectors $\mathbf{R}_I$, of a supercell discussed thoroughly in the next sections. A perturbation calculation then gives:
\begin{eqnarray}
\mathbf{v}_{\mathbf{k} n, \alpha} = \frac{1}{\hbar} \frac{d \epsilon_{\mathbf{k} n}}{d \mathbf{k}_{\alpha}} = \frac{1}{\hbar}
\langle n\mathbf{k} | \frac{d\mathbf{H}_{\mathbf{k}}}{d \mathbf{k}_{\alpha}} - \epsilon_{\mathbf{k} n} \frac{d \mathbf{S}_{\mathbf{k}}}{d \mathbf{k}_{\alpha}} | n\mathbf{k} \rangle \,,
\label{eq:velocityPerturbationTheory}
\end{eqnarray}
where $\epsilon_{\mathbf{k} n}$ and $| n\mathbf{k} \rangle$ are the energy and Bloch state of band $n$ at wave vector $\mathbf{k}$. We hereby avoid finite difference errors from crossing bands when the bands are not sorted correctly. 

\subsection{Phonons\label{Sec:Phonons}}
The phonon polarization vectors ($\mathbf{e}_\mathbf{q}^\lambda$) and energies $\hbar\omega_{\mathbf{q},\lambda}$ are obtained as solutions to the equation
\begin{eqnarray}
\mathbf{D}(\mathbf{q}) \mathbf{e}_\mathbf{q}^\lambda = \omega_{\mathbf{q},\lambda}^2\mathbf{e}_\mathbf{q}^\lambda \,, \label{eq:phonon-eigenvalue-problem}
\end{eqnarray}
where $\mathbf{q}$ is the phonon momentum and $\lambda$ labels the phonon branch (band) index and $\mathbf{D}(\mathbf{q})$ is the Fourier transformed dynamical matrix. We initially compute the real-space dynamical matrix from a standard finite difference approach\cite{frederiksen_inelastic_2007}, where the elements are given by
\begin{eqnarray}
D_{i\mu j\nu} &=& \frac{1}{\sqrt{M_iM_j}}\frac{\partial^2 E_{\rm tot}(\mathbf{R})}{\partial \mathbf{x}_{i\nu} \partial \mathbf{x}_{j\mu}} \nonumber \\
&\approx& \frac{1}{\sqrt{M_iM_j}} \frac{F_{j\mu}(+\Delta \mathbf{x}_{i\nu}) - F_{j\mu}(-\Delta \mathbf{x}_{i\nu})}{2\Delta \mathbf{x}_{i\nu}} \,,
\end{eqnarray}
where $i,j$ are atom indices and $\mu,\nu$ denote cartesian directions. $E_{\rm tot}(\mathbf{R})$ is the total energy written as a function of all atomic coordinates. The force $F_{j\mu}(+\Delta \mathbf{x}_{i\nu})$ is acting on atom $j$ in direction $\mu$ when atom $i$ is displaced by $\Delta \mathbf{x}_{i\nu}$ in direction $\nu$. The approximate equality sign indicates the approximation inherent in the first order finite difference method. 

\begin{figure}[!htbp]%[!t]%[!htbp]%
\centering
{\includegraphics[width=0.7\linewidth]{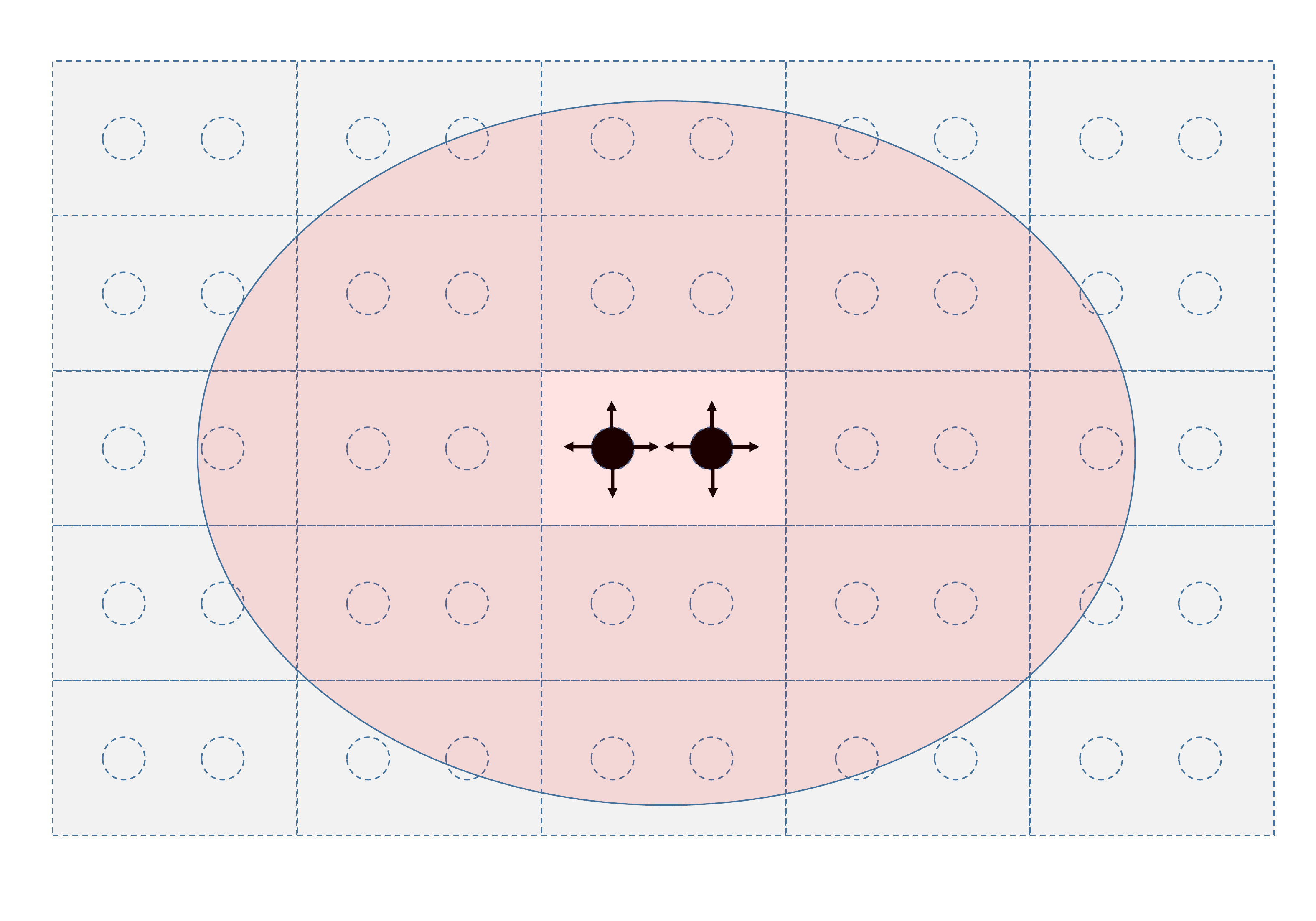}}%width=0.99\linewidth%width=2.5in
\caption{Schematic illustration of the supercell method. A unit cell is repeated 5x5 times. Only the atoms in the central unit cell are being displaced, while forces are evaluated on all the atoms in the supercell. Outside a certain range (elipsoidal area) the forces are zero. }
\label{fig:super-cell}
\end{figure}

When computing the forces, we first construct a supercell by repeating the unit cell $(n_A, n_B, n_C)$ times along the directions of the primitive lattice vectors. Secondly we perform the finite difference derivative by calculating the forces in the entire supercell, while only displacing the atom in the central unit cell, as schematically shown in Fig. \ref{fig:super-cell}.

The normalized phonon eigen modes in Eq.~\eqref{eq:phonon-eigenvalue-problem} are dimensionless. The transformation to modes with physical dimension is
$\mathbf{u}_\mathbf{q}^\lambda = l_q \mathbf{e}_\mathbf{q}^\lambda$, where the characteristic length is calculated from the polarization vectors and the diagonal mass matrix, $\mathbf{m}$:
\begin{eqnarray}
l^{\lambda}_{\mathbf{q}} &=&\sqrt{\frac{\hbar}{2\omega_{\lambda} \,\mathbf{e}^{\lambda \dagger}_{\mathbf{q}}\cdot \mathbf{m}\cdot  \mathbf{e}^{\lambda}_{\mathbf{q}}}} \,.
\end{eqnarray}

\subsection{Bulk electron-phonon coupling\label{Sec:BulkEph}}
% Bulk e-ph method.
%In a periodic cell
We here provide a simple derivation of the bulk electron-phonon coupling in the case of a localized basis by applying the periodicity of the problem.

We want to calculate the coupling matrix element between Bloch states $|n\mathbf{k}\rangle$ and $|n'\mathbf{k}'\rangle$ due to a phonon with momentum $\mathbf{q}$ and branch index $\lambda$ perturbing the Hamiltonian:
\begin{eqnarray}
g_{\mathbf{k}\mathbf{k}'\mathbf{q}}^{\lambda n n'} = \langle n'\mathbf{k}' | \delta \hat{H}_{\mathbf{q}\lambda} | n\mathbf{k}\rangle \,.
\label{eq:M-definition}
\end{eqnarray}

The perturbation to the Hamiltonian can be expressed as:
\begin{eqnarray}
 \delta \hat{H}_{\mathbf{q}\lambda} = l_\mathbf{q}^\lambda \sum_\alpha \sum_I \frac{\partial \hat{H}}{\partial \mathbf{x}_{I,\alpha}} \mathbf{e}_{\mathbf{q},I,\alpha}^\lambda \,,
\label{eq:dH-definition}
\end{eqnarray}
where the $I$-sum runs over the periodic unit cells and the $\alpha$-sum runs over the spatial degrees of freedom (atom index and Cartesian direction) within each cell. Using the Bloch periodicity of the phonon polarization vector we can rewrite this as
\begin{eqnarray}
 \delta \hat{H}_{\mathbf{q}\lambda} = l_\mathbf{q}^\lambda \sum_\alpha \mathbf{e}_{\mathbf{q},\alpha}^\lambda \sum_I \frac{\partial \hat{H}}{\partial \mathbf{x}_{I,\alpha}} e^{i\mathbf{q}\cdot\mathbf{R}_I} \,,
\label{eq:dH-definition2}
\end{eqnarray}
where now $\mathbf{e}_{\mathbf{q},\alpha}^\lambda$ is components of the polarization vector in the unit cell with index '0' (the reference cell). The unit cell with index $I$ is displaced from the reference cell by lattice vector $\mathbf{R}_I$.

We evaluate the derivative of the Hamiltonian in a similar manner as the dynamical matrix described above. A unit cell is repeated to form a supercell, but only the atoms in the central unit cell are displaced. The terms that contribute to the derivative is the (local) effective potential $V_{local}(\mathbf{r})$ and the non-local (NL) Kleinmann-Bylander term $V_{NL}$. Details of the derivative of the Hamiltonian are given in Appendix \ref{HamiltonianDerivative-appendix}. 

The electronic Bloch-states are expressed as
\begin{eqnarray}
|n\mathbf{k}\rangle = \frac{1}{\sqrt{N}}\sum_I\sum_\mu c_{n,\mathbf{k}}^\mu e^{i\mathbf{k}\cdot\mathbf{r}}|\phi_\mu;\mathbf{R}_I\rangle \,, \label{eq:Bloch-state}
\end{eqnarray}
where the $I$-sum runs over $N$ unit cells in the macroscopic system, $\mu$ labels the basis orbitals within a single unit cell, $c_{n,\mathbf{k}}^\mu$ are expansion coefficients, and $|\phi_\mu;\mathbf{R}_I\rangle$ is the $\mu$'th basis orbital in the unit cell displaced from the reference cell by the lattice vector $\mathbf{R}_I$. Inserting \eqref{eq:dH-definition2} and \eqref{eq:Bloch-state} in \eqref{eq:M-definition} we get:
\begin{eqnarray}
g_{\mathbf{k}\mathbf{k}'\mathbf{q}}^{\lambda n n'} &=& \frac{l_\mathbf{q}^\lambda}{N}\sum_{IJK}\sum_{\mu\nu}\sum_\alpha  e^{-i\mathbf{k}'\cdot\mathbf{R}_J} e^{i\mathbf{k}\cdot\mathbf{R}_K} 
(c_{n',\mathbf{k}'}^\nu)^* c_{n,\mathbf{k}}^\mu \nonumber \\
 &\times& \mathbf{e}_{\mathbf{q},\alpha}^\lambda \langle \phi_\nu; \mathbf{R}_J | \frac{\partial \hat{H}}{\partial \mathbf{x}_{I,\alpha}} e^{i\mathbf{q}\cdot\mathbf{R}_I} | \phi_\mu; \mathbf{R}_K \rangle  \,. \label{eq:M1}
\end{eqnarray}
Due to the periodicity of the system, the derivative of the Hamiltonian matrix with respect to atom positions in $\mathbf{R}_I$, can be shifted as follows:
\begin{eqnarray}
\langle \phi_\nu; \mathbf{R}_I+\mathbf{R}_m | \frac{\partial \hat{H}}{\partial \mathbf{x}_{I,\alpha}} | \phi_\mu; \mathbf{R}_I+\mathbf{R}_l \rangle
= \langle \phi_\nu; \mathbf{R}_m | \frac{\partial \hat{H}}{\partial \mathbf{x}_{0,\alpha}} | \phi_\mu; \mathbf{R}_l \rangle \,. \nonumber
\end{eqnarray}
where we defined the relative vectors $\mathbf{R}_{m/l}$ connecting cell $K,J$ to the cell $I$.
As for the force derivative calculation described in Section \ref{Sec:Phonons}, the derivative of the Hamiltonian will also be non-zero in a region around the atoms being displaced. The $J,K$-sums in Eq.~\eqref{eq:M1} which runs over all cells in the macroscopic sample can be limited to the cells included in the supercell calculation of the $\partial \hat{H}/\partial \mathbf{x}_{0,\alpha}$. We thus replace the $J,K$-sums with sums over {\it neighbouring} cells $m,l$ relative to $I$ in Eq.~\eqref{eq:M1}:
\begin{eqnarray}
g_{\mathbf{k}\mathbf{k}'\mathbf{q}}^{\lambda n n'} &=& \frac{l_\mathbf{q}^\lambda}{N}\sum_{Iml}\sum_{\mu\nu}\sum_\alpha  e^{-i\mathbf{k}'\cdot(\mathbf{R}_I+\mathbf{R}_m)} e^{i\mathbf{k}\cdot(\mathbf{R}_I+\mathbf{R}_l)} e^{i\mathbf{q}\cdot\mathbf{R}_I}
\nonumber \\
 &\times& (c_{n',\mathbf{k}'}^\nu)^* c_{n,\mathbf{k}}^\mu \mathbf{e}_{\mathbf{q},\alpha}^\lambda \langle \phi_\nu; \mathbf{R}_m | \frac{\partial \hat{H}}{\partial \mathbf{x}_{0,\alpha}} | \phi_\mu; \mathbf{R}_l \rangle  \,, \label{eq:M2}
\end{eqnarray}
where the derivative of the Hamiltonian is only carried out for spatial degrees of freedom ($\alpha$) in the reference unit cell. 
The $I$-sum can now be carried out to simply give a factor of $\sum_I e^{i (\mathbf{k}-\mathbf{k}'+\mathbf{q})\cdot\mathbf{R}_I}=N\,\delta_{\mathbf{k}',\mathbf{k}+\mathbf{q}}$ which enforces momentum conservation. 
Defining $g_{\mathbf{k}\mathbf{k}'\mathbf{q}}^{\lambda n n'}=g_{\mathbf{k}\mathbf{q}}^{\lambda n n'} \delta_{\mathbf{k}',\mathbf{k}+\mathbf{q}}$, Eq.~\eqref{eq:M2} is then simplified to the final expression for the bulk electron-phonon coupling in a supercell setup:
%Due to the periodicity of the system, the $I$-sum can be carried out to simply give a factor of $N$ (). Keeping only $I=0$ terms,  Eq.~\eqref{eq:M1} then simplifies to 
%\begin{eqnarray}
%g_{\mathbf{k}\mathbf{q}}^{\lambda n n'} &=& l_\mathbf{q}^\lambda\sum_{JK}\sum_{\mu\nu}\sum_\alpha  e^{-i(\mathbf{k}+\mathbf{q})\cdot\mathbf{R}_J} e^{i\mathbf{k}\cdot\mathbf{R}_K} 
%(c_{n',\mathbf{k}+\mathbf{q}}^\nu)^* c_{n,\mathbf{k}}^\mu \nonumber \\
% &\times& \mathbf{e}_{\mathbf{q},\alpha}^\lambda \langle \phi_\nu; \mathbf{R}_J | \frac{\partial \hat{H}}{\partial \mathbf{x}_{0,\alpha}} | \phi_\mu; \mathbf{R}_K \rangle,  \label{eq:M2}
%\end{eqnarray}
\begin{eqnarray}
g_{\mathbf{k}\mathbf{q}}^{\lambda n n'} &=& l_\mathbf{q}^\lambda\sum_{ml}\sum_{\mu\nu}\sum_\alpha  e^{i\mathbf{k}\cdot\mathbf{R}_l-i(\mathbf{k}+\mathbf{q})\cdot\mathbf{R}_m} 
(c_{n',\mathbf{k}+\mathbf{q}}^\nu)^* c_{n,\mathbf{k}}^\mu \nonumber \\
 &\times& \mathbf{e}_{\mathbf{q},\alpha}^\lambda \langle \phi_\nu; \mathbf{R}_m | \frac{\partial \hat{H}}{\partial \mathbf{x}_{0,\alpha}} | \phi_\mu; \mathbf{R}_l \rangle \,.  \label{eq:M3}
\end{eqnarray}

In summary, Eq.~\eqref{eq:M3} provides a procedure for calculating the bulk electron-phonon coupling in any localized basis setup:
One has to evaluate the finite differences of a supercell Hamiltonian where atoms in the center cell are displaced and a summation over unit cells is performed with corresponding phase factors.\cite{Note2}
%with summation over unit cells
% in a given code.
%When applying the expression in a given code one has to choose sign of the phases that are consistent with direction applied in the phonon and electronic structure calculations.

\section{Simulations and results\label{Sec:Simulations}}
The simulations were performed using the ATK DFT code with the PBE-GGA
functional for exchange-correlation in the cases of graphene and silicene, and LDA in the case of MoS$_2$.
In all cases we use a Double-Zeta-Polarized (DZP) basis-set. The real-space grid cutoff
was $110\,\mathrm{Ha}$. The geometries were relaxed until all forces were smaller than $0.001\,\mathrm{eV/\AA}$, and $51\times51$ in-plane $k$-points were used in the electronic structure calculations.
A vacuum gap of $30\,\mathrm{\AA}$ was used in the direction normal to the material plane and Dirichlet boundary conditions was used in the Poisson equation for this direction.
The bulk electron-phonon interaction and phonon dispersion was obtained from a $11\times11$ supercell calculation in the case of graphene and a $9\times9$ supercell for silicene and MoS$_2$. The delta-functions in Eq.~\eqref{eqn:FGRtransitionRate} were numerically represented by Lorentzians with a broadening of $\gamma=3\,\mathrm{meV}$.
%The transmission data have subsequently been interpolated.\cite{Falkenberg2015}

\subsection{Bandstructures}
\begin{figure}[!htbp]%[!t]%[!htbp]%
\centering
{\includegraphics[width=0.99\linewidth]{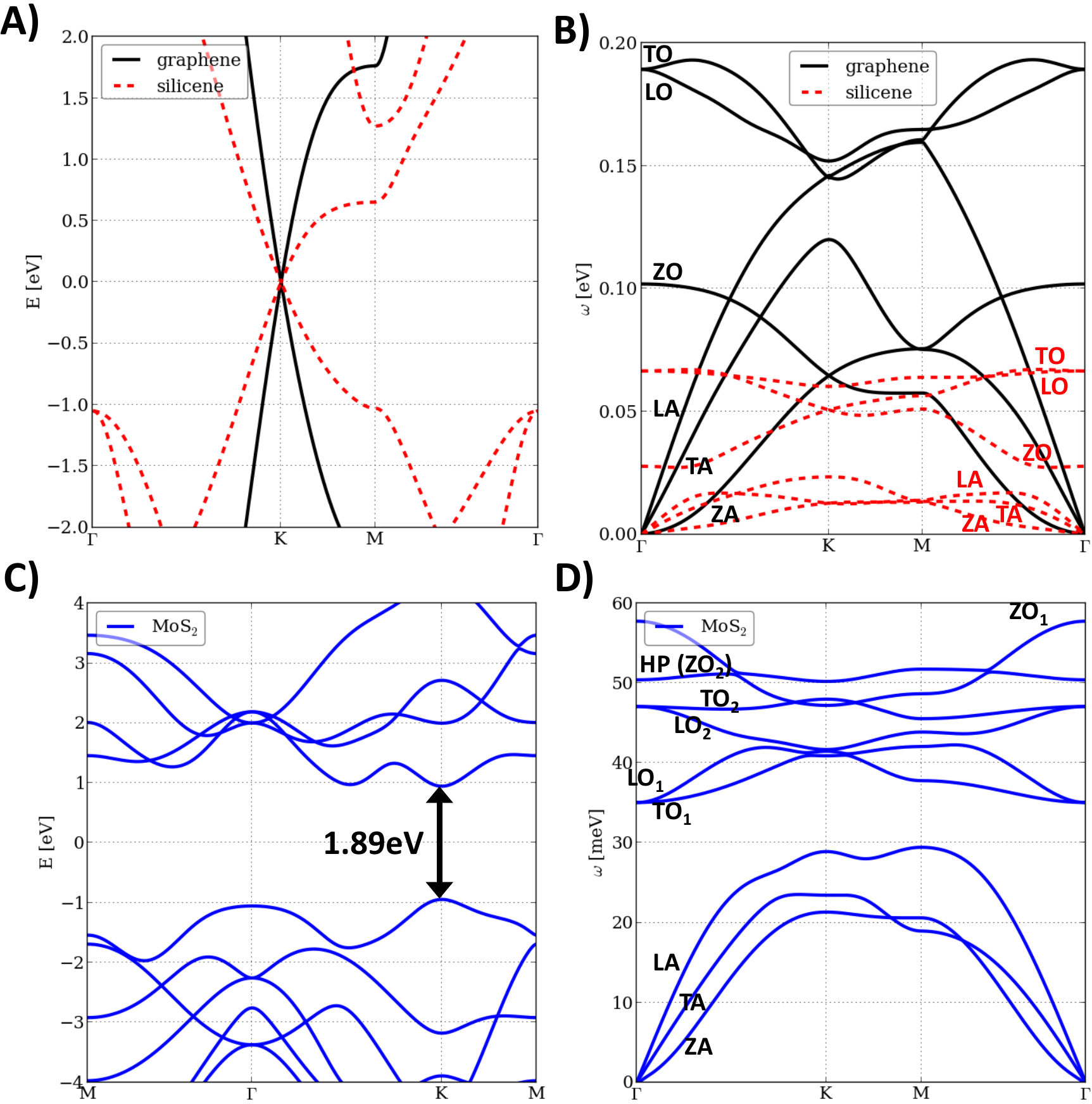}}%width=0.99\linewidth%width=2.5in
\caption{(Color online) Electron bandstructure of A) graphene and silicene and C) MoS$_2$ and phonon dispersion of  B) graphene and silicene and D) MoS$_2$.}
\label{fig:Bandstructures}
\end{figure}
One obtains linear valence and conduction bands near the Dirac point, $K$, in both graphene and silicene, as shown by the band structures in Fig.~\ref{fig:Bandstructures}A. We obtain Fermi velocities of $0.9\times 10^{6}\,\mathrm{m/s}$ and $0.57\times 10^{6}\,\mathrm{m/s}$ of graphene and silicene, respectively. Both materials have six phonon branches. The three acoustic modes will dominate the low temperature scattering where two modes (LA, TA) have a linear q-dependence and the third out-of-plane acoustic (ZA) mode has a q$^2$-dependence near the Brillouin zone center\cite{gu_first-principles_2015}, see Fig.~\ref{fig:Bandstructures}B. We obtain sound velocities of $20.4(12.6)\times 10^{3}\,\mathrm{m/s}$ for the LA(TA) mode of graphene and $9.1(6.1)\times 10^{3}\,\mathrm{m/s}$ for the LA(TA) mode of silicene.
MoS$_2$ is found to be a direct-gap semiconductor\cite{mak_atomically_2010,splendiani_emerging_2010} with a band gap of $1.89\,\mathrm{eV}$, see Fig.~\ref{fig:Bandstructures}C. The electron- and phonon bandstructures, in Fig.~\ref{fig:Bandstructures}C-\ref{fig:Bandstructures}D, are consistent with previous theoretical results.\cite{kaasbjerg_phonon-limited_2012}
MoS$_2$ has three acoustic and six optical branches. The three acoustic branches are the in-plane
longitudinal acoustic (LA), the transverse acoustic (TA) as well as the out-of-plane acoustic (ZA) modes.
We obtain sound velocities of $6.6(4.2)\times 10^{3}\,\mathrm{m/s}$ for the LA(TA) mode of MoS$_2$.
The six optical branches are two in-plane longitudinal optical (LO$_1$, LO$_2$), two in-plane transverse optical (TO$_1$, TO$_2$) and two out-of-plane optical (ZO$_1$, ZO$_2$) modes. The two lowest optical branches (LO$_1$, TO$_1$) are nonpolar modes which do not couple to the charge carriers.
The next two branches (LO$_2$, TO$_2$) are polar optical modes where the Mo and S atoms vibrate in counterphase. The dispersionless out-of-plane mode, ZO$_2$, is also called the homopolar mode. It is characteristic of layered structures and is related to fluctuations in the layer thickness\cite{fivaz_mobility_1967}.
For further discussion of phonon modes in MoS$_2$ we refer the reader to Refs. \onlinecite{zhang_phonon_2015,molina-sanchez_phonons_2011}.

\subsection{Bulk electron-phonon coupling}
We now turn to the calculated bulk electron-phonon coupling.
It is common practice to plot the scattering matrix element for a fixed $\mathbf{k}$-point as a function of $\mathbf{q}$.
In this way one can visualize the detailed suppression of the scattering, as opposed to a constant or linear-in-$q$ deformation potential, depending on the symmetry of the involved phonon and electron states.
The units are converted to eV/\AA, often used for extracted deformation potentials from experiments, by dividing by the characteristic length prefactor, $l^{\lambda}_{\mathbf{q}}$, in Eq.~\eqref{eq:M3}.
\begin{figure}[!htbp]%[!t]%[!htbp]%
\centering
{\includegraphics[width=0.99\linewidth]{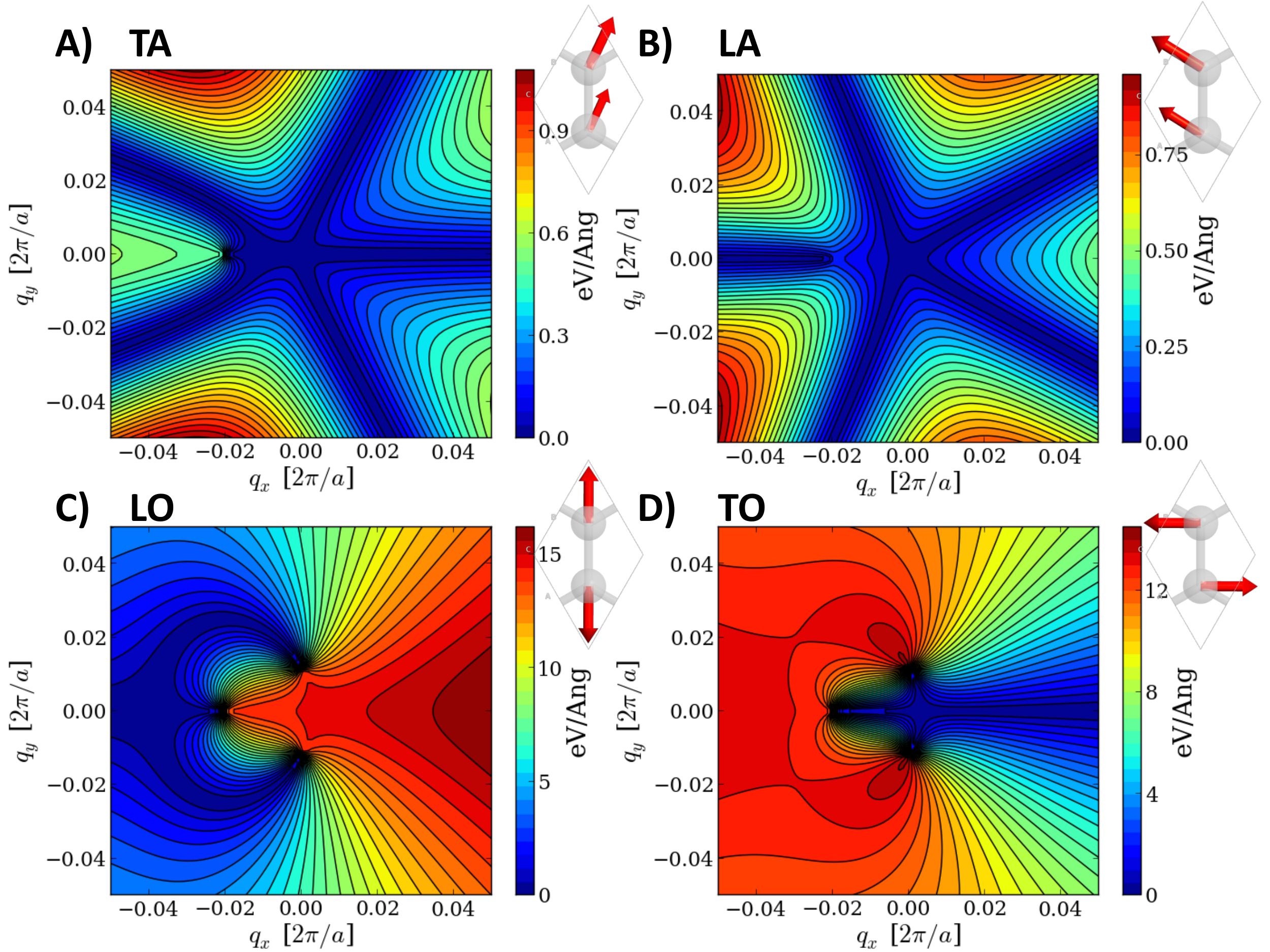}}%width=0.99\linewidth%width=2.5in
\caption{(Color online) Bulk electron-phonon coupling in graphene for the four modes with nonzero coupling. The interaction is illustrated as a function of phonon $\mathbf{q}$-vector at a $\mathbf{k}$-point shifted $300\,\mathrm{meV}$ from the Dirac $K$-point towards the $\Gamma$-point. We refer the reader to 
Ref.~\onlinecite{kaasbjerg_unraveling_2012} for a detailed discussion of the interpretation of the plots for the TA and LA modes. The scattering rate is obtained as integrals around the constant energy circles satisfying $\varepsilon_{\mathbf{k'}}=\varepsilon_{\mathbf{k}}\pm\hbar\omega$. Inserts: phonon modes are vizualized by arrows indicating the atomic displacements.}
\label{fig:graphene_eph}%kaasbjerg \textit{et al.}\cite{kaasbjerg_unraveling_2012}
\end{figure}
In Fig.~\ref{fig:graphene_eph} we illustrate the $\mathbf{q}$-variation of the bulk electron-phonon interaction obtained for the four modes coupling with electrons in graphene.
The atomic motion of the considered mode is illustrated as an insert in the upper right corner, by arrows indicating the atomic displacement.
The interaction is obtained around $300\,\mathrm{meV}$ from the Dirac $K$-point, and the interaction seems slightly higher for the acoustic modes (between $10-20$\%) and very close in magnitude for the optical modes (within $10$\%), but in both cases with the same symmetry as two previously published results.\cite{kaasbjerg_unraveling_2012,park_electronphonon_2014} The maximal values for the TO and LO modes are in energy units approximately $0.4-0.5\,$eV for comparison.
Importantly, we see that the coupling elements are highly anisotropic. In the case of acoustic modes we see that back-scattering ($q_x<0,q_y\approx 0$) is suppressed for the LA mode, while the situation is reversed for the TA mode where forward scattering is suppressed.
In addition, other directions with complete suppression also appear and the $q$-dependence is highly nontrivial. In general, the anisotropy and scattering suppression of the bulk electron-phonon coupling is determined by the combined symmetry of both phonon and electronic states.

%The calculated electron-phonon coupling matrix elements are in agreement with plane-wave calculations for graphene.\cite{kaasbjerg_unraveling_2012} 
\begin{figure}[!htbp]%[!t]%[!htbp]%
\centering
{\includegraphics[width=0.99\linewidth]{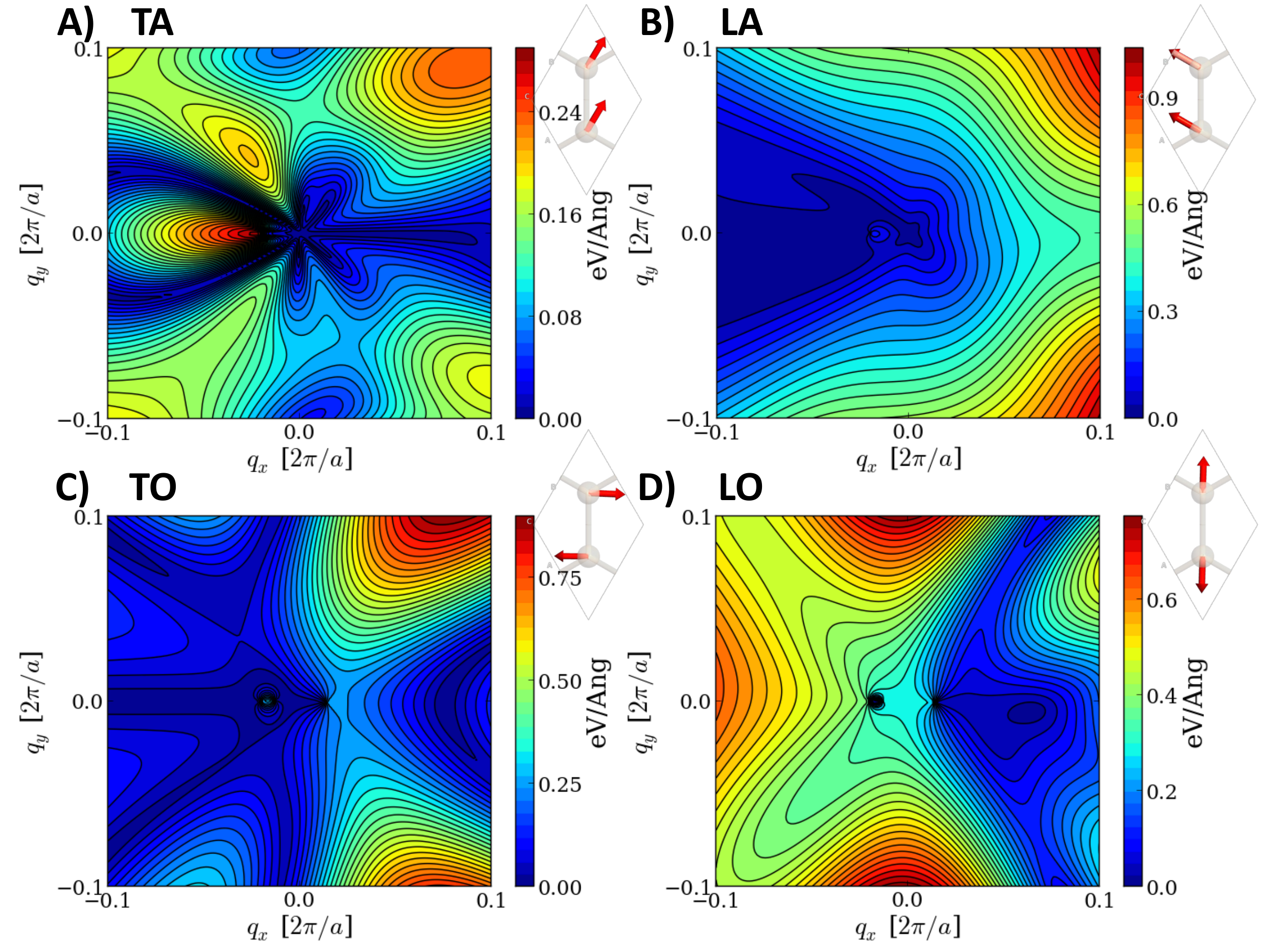}}\\%width=0.99\linewidth%width=2.5in
{\includegraphics[width=0.99\linewidth]{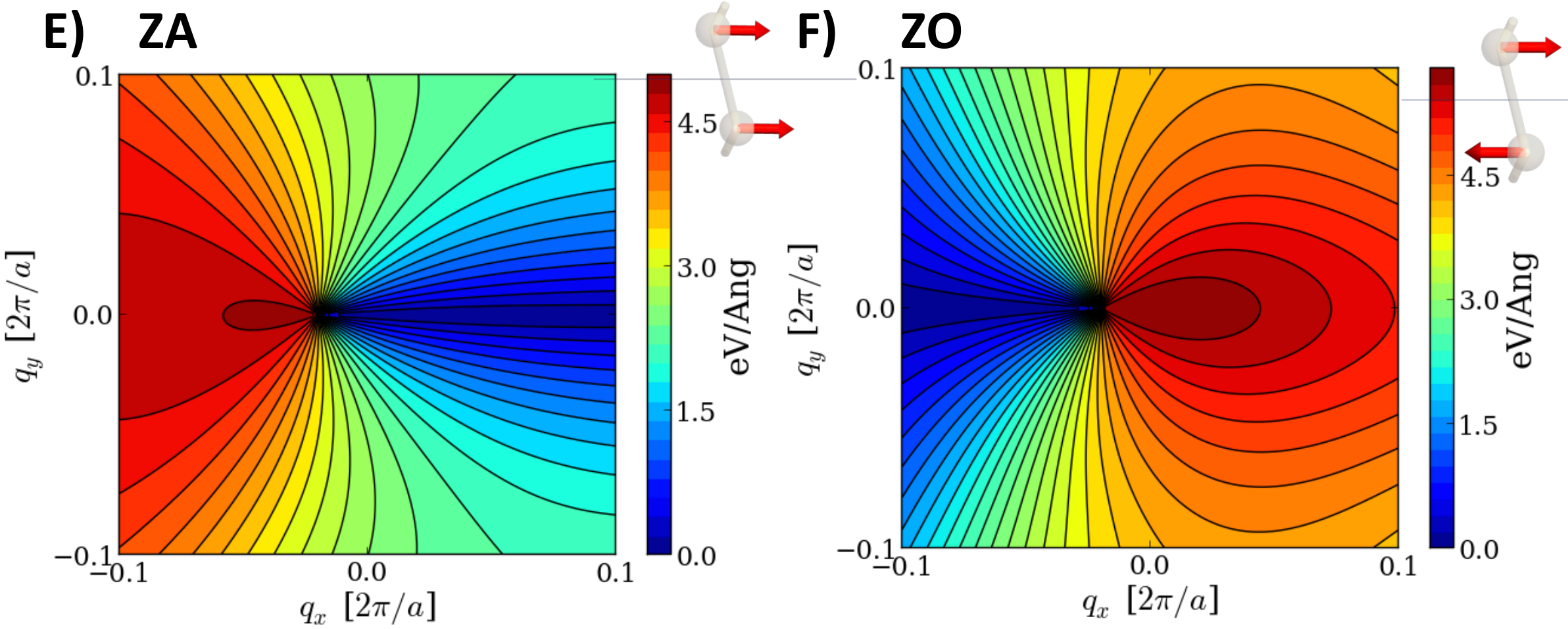}}%width=0.99\linewidth%width=2.5in
\caption{(Color online) Bulk electron-phonon coupling in silicene. Unlike graphene also the out-of-plane modes (ZA, ZO) couple significantly with electrons. The interaction is illustrated as a function of phonon $\mathbf{q}$-vector at a $\mathbf{k}$-point shifted $110\,\mathrm{meV}$ from the Dirac $K$-point towards the $\Gamma$-point. The scattering rate is obtained as an integral around a constant energy circle satisfying $\varepsilon_{\mathbf{k'}}=\varepsilon_{\mathbf{k}}\pm\hbar\omega$. Inserts: phonon modes are vizualized by arrows indicating the atomic displacements.}
\label{fig:silicene_eph}
\end{figure}
Unlike graphene, the carriers in silicene display a strong interaction with the ZA mode,\cite{li_intrinsic_2013} see Fig.~\ref{fig:silicene_eph}. This is related to the buckling of the silicene sheet, where one basis atom is displaced approximately $0.44\,\mathrm{\AA}$ out-of-plane hence breaking the planar symmetry.
Otherwise we see that the situation for the acoustic modes is similar to that of graphene.
Back-scattering is suppressed for the LA mode, while the situation is reversed for the TA mode where forward scattering is suppressed. Again, other directions with complete suppression of the scattering also appear.
%Fig.1E illustrates the scattering rate for k-points along the $\Gamma$-K direction moving away from a Dirac point.

% MoS2
It is important to realize that the reduction of the symmetry of the lattice leads to additional scattering mechanisms and modes interacting with the charge carriers.
This was the case for silicene where there was no planar symmetry, and it is even more so for the case of MoS$_2$ which lacks inversion symmetry.
In the last mentioned case, all kinds of electron-phonon coupling may take place; like different orders of deformation potential, piezoelectric and Fr\"{o}hlich couplings.
This makes a strong case for a fully numerical solution of the BTE together with a first-principles method to evaluate the bulk electron-phonon interactions. We avoid having a high number of free parameters in an analytical model and instead every type of interaction is directly taken care of.
\begin{figure}[!htbp]%[!t]%[!htbp]%
\centering
{\includegraphics[width=0.99\linewidth]{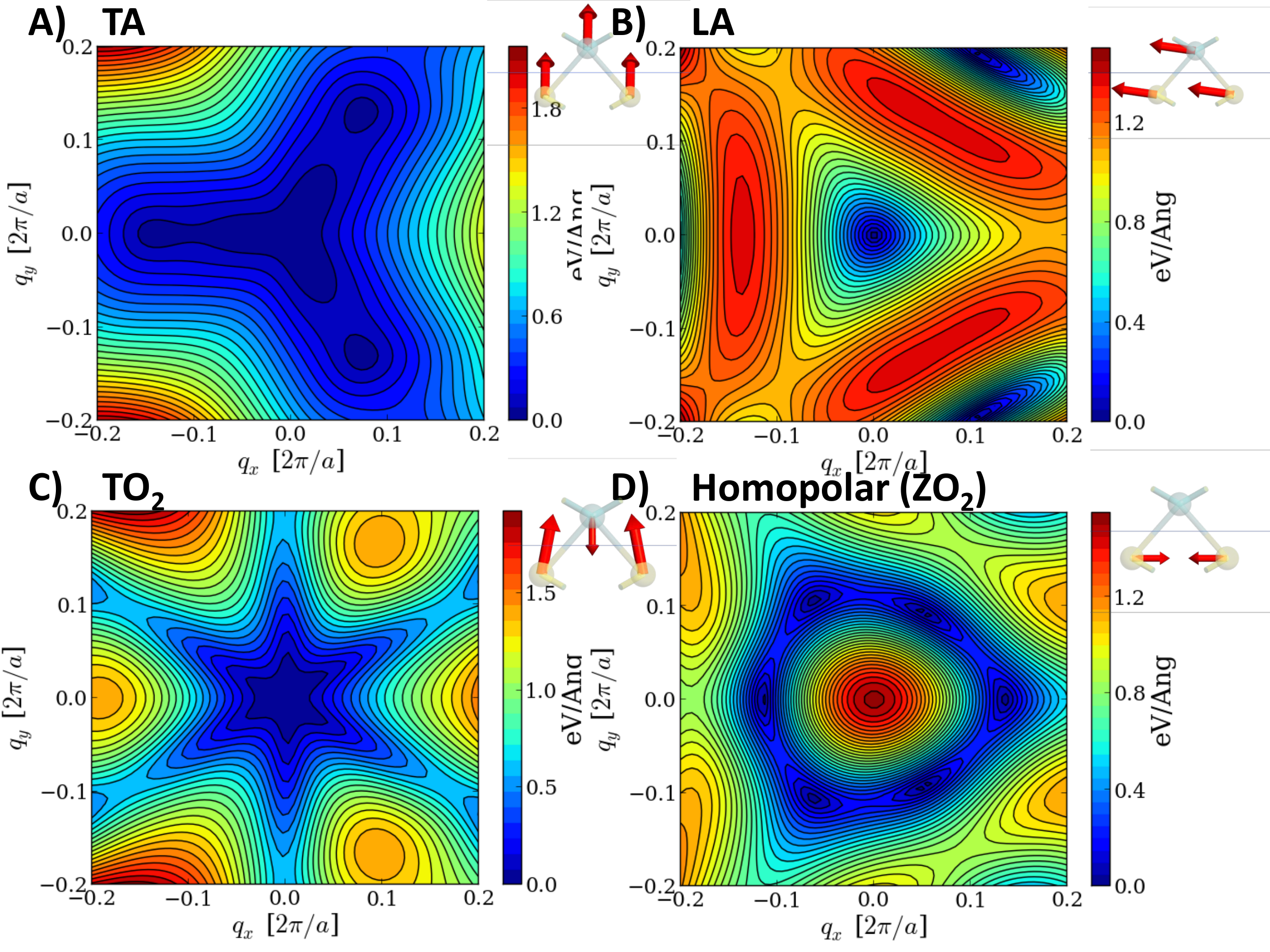}}\\%width=0.99\linewidth%width=2.5in
{\includegraphics[width=0.499\linewidth]{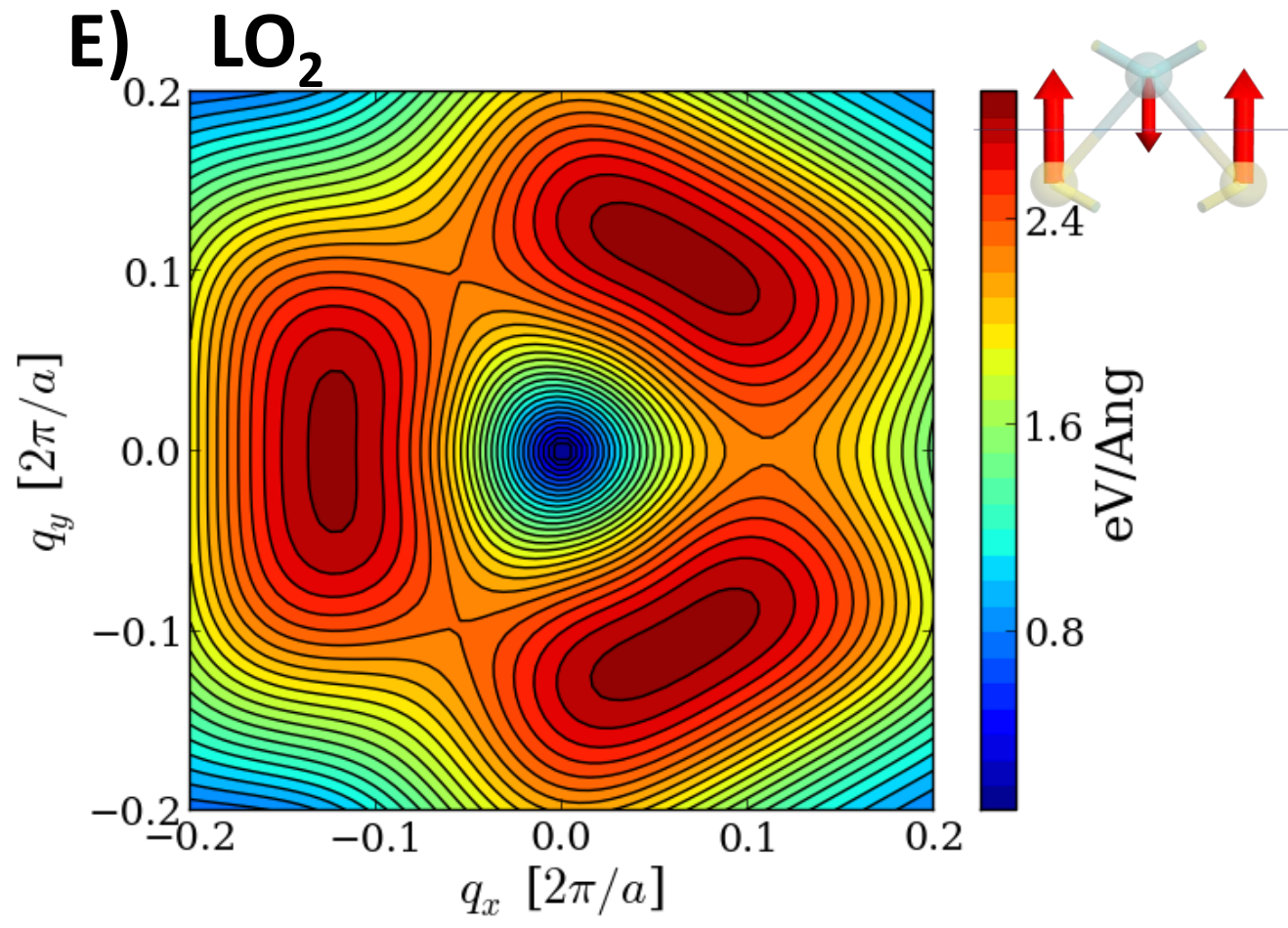}}%width=0.99\linewidth%width=2.5in
\caption{(Color online) Bulk electron-phonon coupling in MoS$_2$. The interaction is illustrated as a function of phonon $\mathbf{q}$-vector at the conduction band minimum $K$-point. Inserts: phonon modes are vizualized by arrows indicating the atomic displacements.}%displayed as an exaggerated overlay of atom positions from small to large radius.
\label{fig:MoS2_eph}
\end{figure}
In Fig.~\ref{fig:MoS2_eph} we show the bulk electron-phonon interaction obtained for the five modes coupling with electrons in MoS$_2$. Unlike graphene and silicene, which are semimetals, MoS$_2$ is a semiconductor and the interaction is evaluated at the conduction band minimum, since only n-doping is relevant in MoS$_2$. Again we find very anisotropic couplings where the symmetry compares well to previously published results.\cite{kaasbjerg_phonon-limited_2012}

The coupling with the TA, LA and TO modes are of the same order of magnitude as in Ref.~\onlinecite{kaasbjerg_phonon-limited_2012}, cf. Fig.~\ref{fig:MoS2_eph}A,B and C, but we obtain a somewhat lower coupling for the homopolar mode (approximately 65\% lower), see Fig.~\ref{fig:MoS2_eph}D. Fig.~\ref{fig:MoS2_eph}E shows the Fr\"{o}hlich interaction for the polar optical LO$_2$ mode. The Fr\"{o}hlich interaction is difficult to converge with respect to supercell size. In the long-wavelength limit this element should increase linearly.\cite{fivaz_mobility_1967,kaasbjerg_phonon-limited_2012} We find that the peaks increase in magnitude by approximately 4\% and move toward $|\mathbf{q}|\rightarrow 0$ as expected if the supercell size is increased from $9\times9$ to $15\times15$.
%, but the remaining couplings are of the same magnitude

\subsection{Scattering rates}
\begin{figure*}[!htbp]%[!t]%[!htbp]%
\centering
{\includegraphics[width=0.99\linewidth]{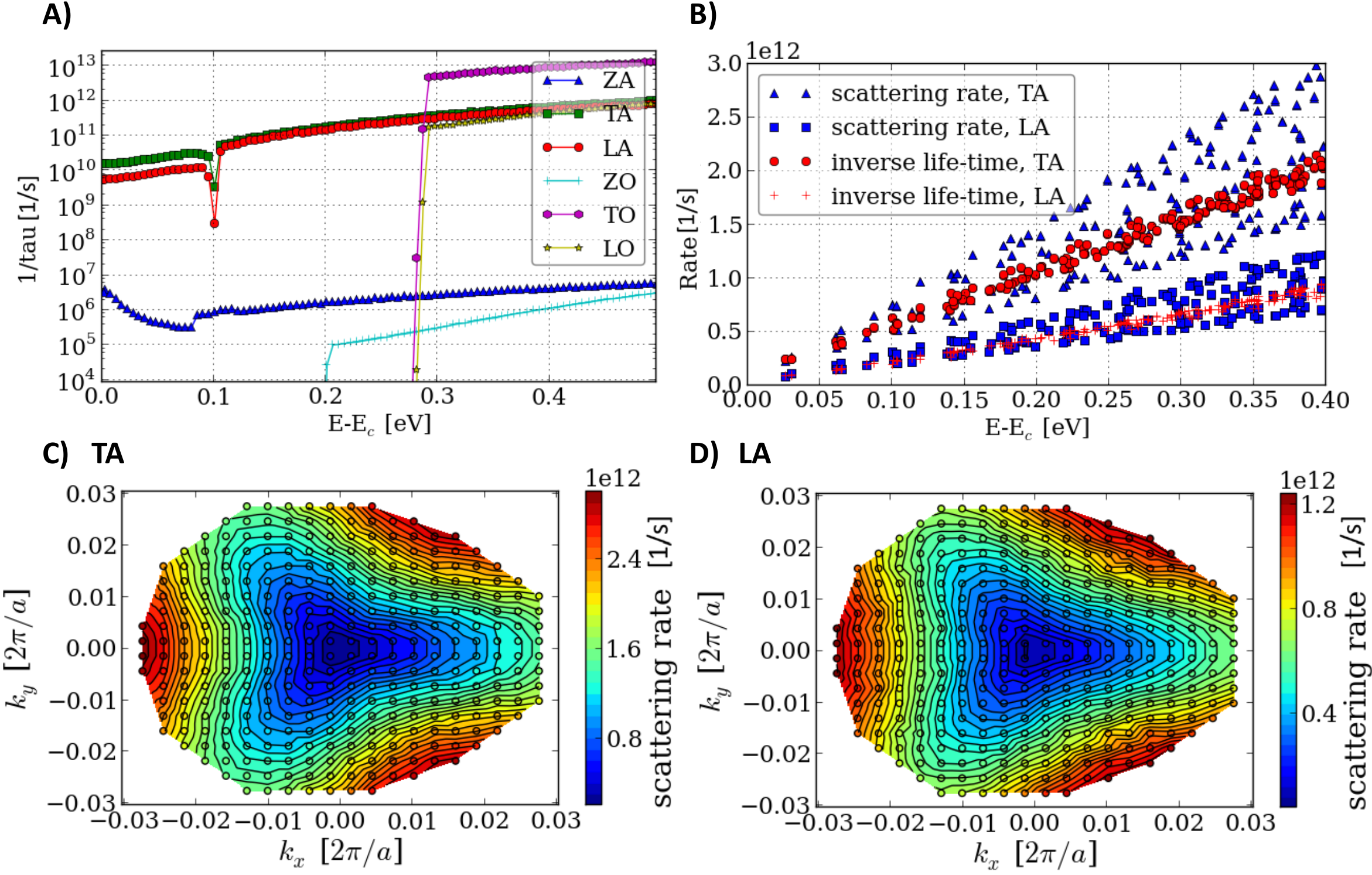}}%width=0.99\linewidth%width=2.5in
\caption{(Color online) Scattering rate as a function of energy/$\mathbf{k}$ for graphene.
A) At a temperature of 5.7\,K and a Fermi-level $\mu_F=100$\,meV along the $\mathbf{k}$-points from $K$ to $\Gamma$ up to 0.5\,eV. 
B) Comparison of the anisotropy in the scattering rate and inverse lifetime at a temperature of 300\,K and a Fermi-level $\mu_F=100$\,meV.
C,D) Two-dimensional scattering rate for the highly anisotropic TA and LA phonon modes. The $\mathbf{k}$-points are relative to the Dirac point, $K$, and the temperature is 300\,K and Fermi-level $\mu_F=100$\,meV. The points illustrate the original mesh up to 0.5\,eV and the contours are obtained from cubic spline interpolation.}
\label{fig:RateGraphene}
\end{figure*}

% Rates for graphene/silicene/MoS$_2$
We obtain the scattering rate by integrating the linearized BTE within the RTA, cf. Eq.~\eqref{eqn:RTA_rate}.
The electron-phonon coupling is evaluated for every $\mathbf{k}$-point up to an energy-cutoff in a $100\times100$ $\mathbf{q}$-mesh. The coupling, energies and velocities were subsequently interpolated to twice this $\mathbf{q}$-space resolution before the BTE was solved.
In Fig.~\ref{fig:RateGraphene} we show the result obtained for graphene.
Below the Bloch-Gr\"uneisen temperature (approximate 57$\sqrt{n}$\,K for graphene with the carrier density measured in units of $10^{12}$cm$^{-2}$), only those phonons with short $\mathbf{q}$ are effectively excited. This manifest itself in the dip in the scattering rate around the Fermi-level.
We clearly see the expected low-temperature Bloch-Gr\"uneisen dips around the Fermi-level (100meV in the present case) and the opening of optical phonon interaction (emission) at $\mu_F+\hbar \omega$, Fig.~\ref{fig:RateGraphene}A. It is illustrative to plot the rate along a single $\mathbf{k}$-line as in Fig.~\ref{fig:RateGraphene}A. However, the full two-dimensional dependency is needed to capture the anisotropy of the scattering rate.
Fig.~\ref{fig:RateGraphene}B illustrate the scattering rate and inverse lifetime of the LA and TA modes found in a full two-dimensional $\mathbf{k}$-mesh. 
The spread of the points illustrate a significant dependence on directions of the scattering rate. This if further highlighted in Fig.~\ref{fig:RateGraphene}C,D where the scattering rate was interpolated to clearly illustrate the anisotropy of especially the LA and TA modes.
Part of the anisotropy originates from the bulk electron-phonon coupling. This is the main contribution to the anisotropy of the inverse lifetime.
However, the anisotropy is further amplified by the transport scattering angle $1-\rm{cos}(\theta)$ in Eq.~\eqref{eqn:RTA_rate}, which is seen by comparing scattering rate and inverse lifetime in Fig.~\ref{fig:RateGraphene}B.

We conclude that the scattering rate depends significantly on the $\mathbf{k}$-space directions. In addition, our implementation gives results for the graphene scattering rate that are consistent with previous theoretical results in the low-temperature Bloch-Gr\"uneisen regime, as well as the high temperature equipartition regime where the scattering rate should depend linearly on energy.\cite{hwang_acoustic_2008,kaasbjerg_phonon-limited_2012}

\begin{figure}[!htbp]%[!t]%[!htbp]%
\centering
{\includegraphics[width=0.99\linewidth]{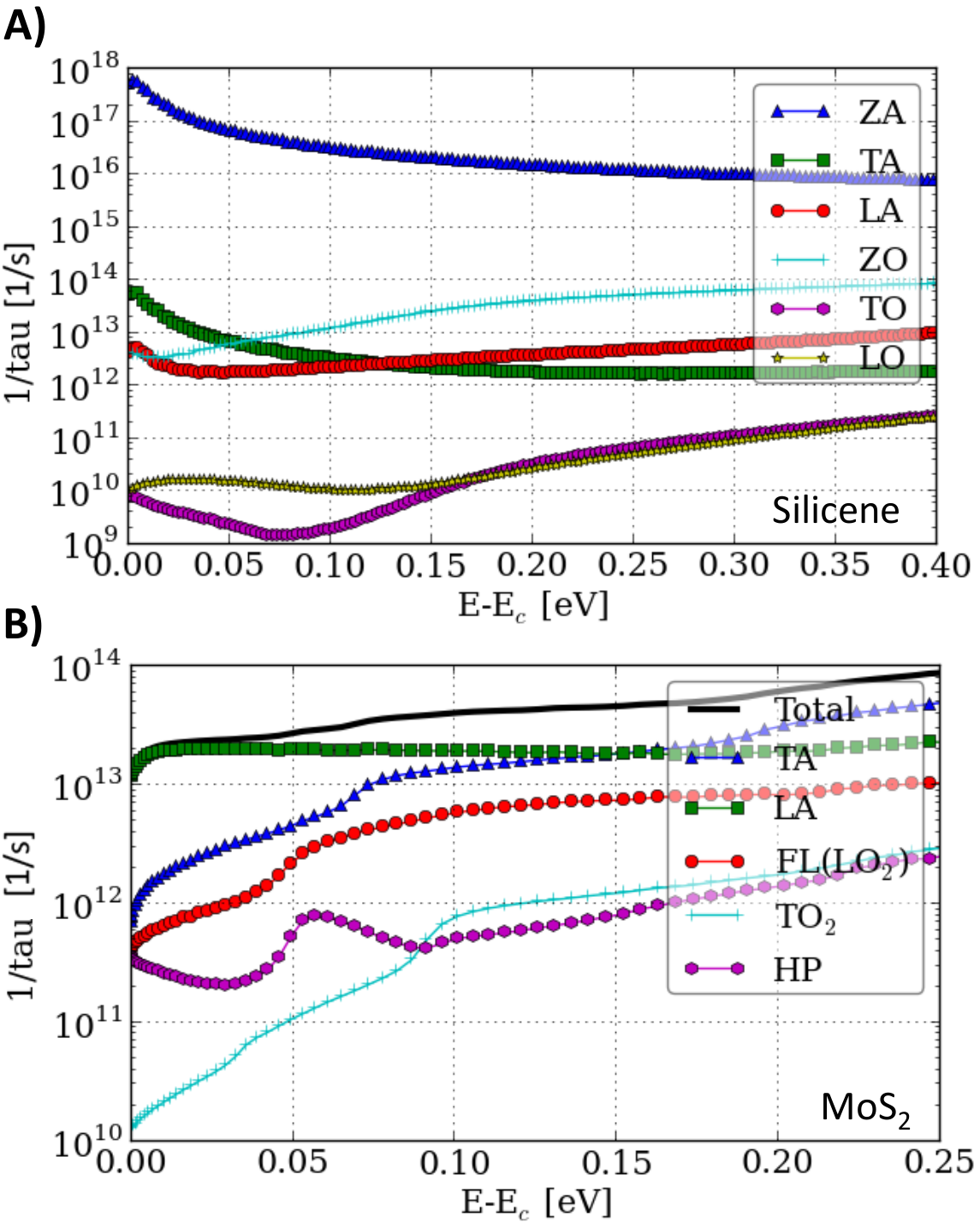}}%width=0.99\linewidth%width=2.5in
\caption{(Color online) Scattering rate as a function of energy/$\mathbf{k}$ for silicene (A) and MoS$_2$ (B) relative to the conduction band minimum $E_c$. In both cases the temperature is 300\,K. The Fermi-level is tuned to $\mu_F=100$\,meV for silicene and 500\,meV below the conduction band minimum for MoS$_2$. We used $\mathbf{k}$-points along a line from $K$ to $\Gamma$ up to 0.5eV for silicene and $K$ to $M$ for MoS$_2$.}
\label{fig:RateSiliceneAndMoS2}
\end{figure}

Previous studies have calculated the lifetime of carriers in silicene.\cite{li_intrinsic_2013}
In Fig.~\ref{fig:RateSiliceneAndMoS2} we show the scattering rate along a single $\mathbf{k}$-line for silicene and MoS$_2$.

We find a significant scattering with the ZA mode for silicene. However, this mode is difficult to describe accurately; The electron-phonon interaction is only partly screened in the present formalism and the linear dispersion of the ZA modes results in a constant density of states which is therefore not able to cutoff long-wavelength interactions. 
%due to the unscreened phonon modes and interactions applied here and the constant phonon density of states obtained from a linear dispersion.
We expect the scattering rate with the out-of-plane ZA mode to be significantly reduced due to interaction with a substrate and a more precise description of screening. Therefore, this interaction will not be included in the mobility calculations. The obtained mobilities should be taken as upper limits and if the ZA scattering is included as is it would reduce the mobility by two orders of magnitude.
For both silicene and MoS$_2$, we find that the rates only changes slightly if the Fermi-level is reduced further.
For MoS$_2$ the scattering rate is almost independent of $\mu_F$ for $\mu_F<E_c$; only the scattering rates for the LO$_2$ and LA modes reduces slightly at high energy.
Unlike the scattering related to microscopic changes in potential (deformation potential or nonpolar optical scattering), the scattering related to the macroscopic electric-field (piezoelectric or Fr\"{o}hlich interaction) will be partially screened by a dielectric environment.
One approach to handle this difference in screening is to partition electron-phonon interaction in real-space into the short-ranged deformation potential contributions and long-ranged piezoelectric and Fr\"{o}hlich interactions and apply the screening individually\cite{kaasbjerg_acoustic_2013}.
However, as shown in Fig.~\ref{fig:RateSiliceneAndMoS2}B the Fr\"{o}hlich interaction (LO$_2$ mode) is not dominating the transport and the piezoelectric coupling is most important at low temperatures\cite{kawamura_phonon-scattering-limited_1992,kaasbjerg_phonon-limited_2012}.

%MoS2: ScatteringRateFinal_ModeCurvesSelection_nq100_refinement1_KM_broadening10meV_Zoom
%300K. All modes except ZA. Fermishift=0.9eV (500meV below Ec). Same at Fermishift=Ec=0.95eV and only changes slightly for Fermishift=0 (EF) where only FL and LA reduces slightly at high energy. That is the rate is almost gate independent for Fermishift<Ec. Above Ec it does change somewhat.
%Silicene: 300K, mu=100meV.
%ScatteringRate_ModeCurvesAll_nq200_refinement1_KG_broadening10meV_T300K

\subsection{Mobility}
\begin{figure*}[!htbp]%[!t]%[!htbp]%
\centering
{\includegraphics[width=0.99\linewidth]{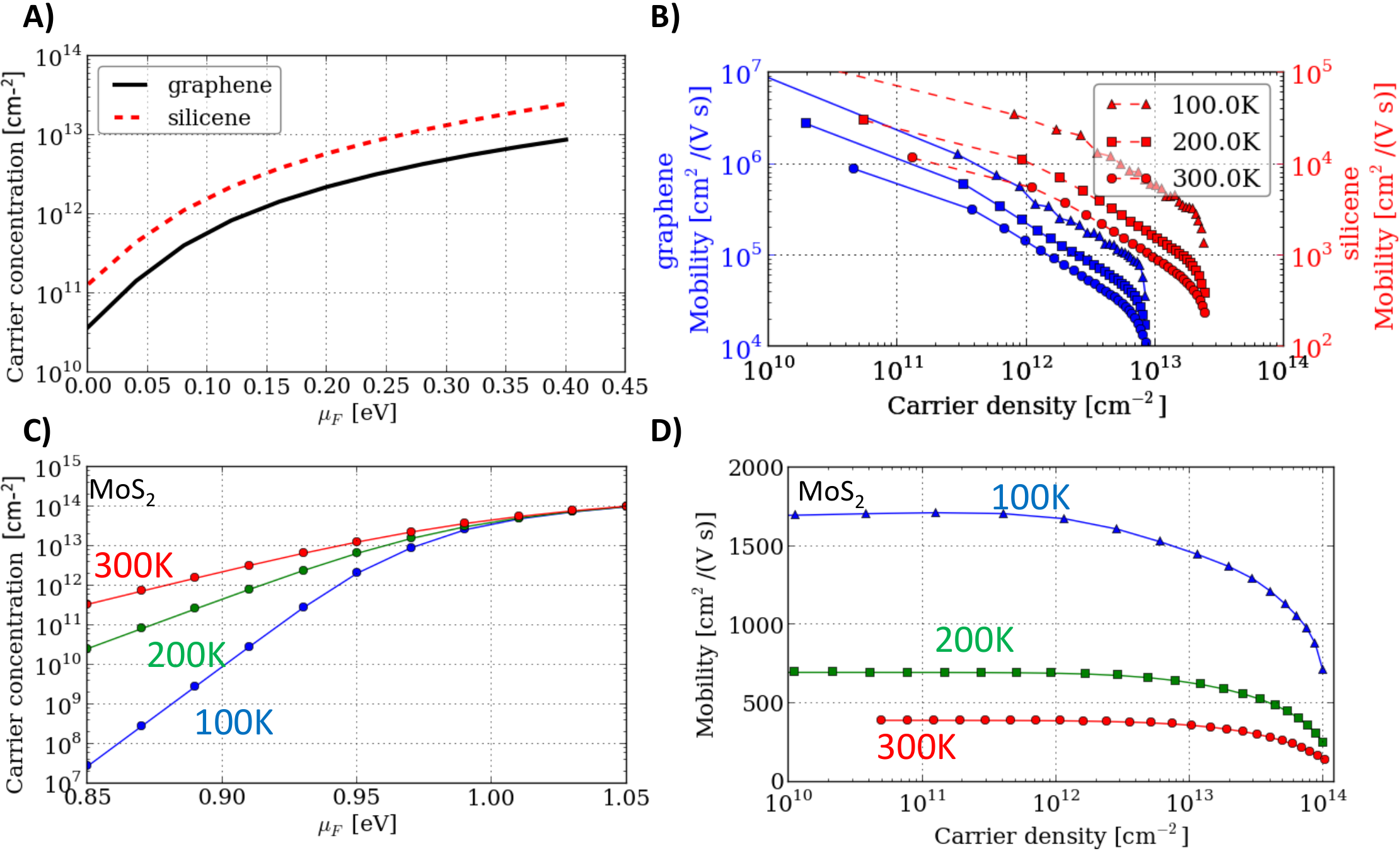}}%width=0.99\linewidth%width=2.5in
\caption{(Color online) Carrier concentration as a function of Fermi-level for graphene and silicene (A) and MoS$_2$ (C). The conduction band edge of MoS$_2$ is at $0.95\,\mathrm{eV}$. The mobility as a function of carrier density is also shown for graphene (B) and silicene (B, inset) and MoS$_2$ (D).}
\label{fig:mobility}
\end{figure*}

% Mobility
In Fig.~\ref{fig:mobility} we illustrate the obtained carrier concentration and mobility\cite{Note3} of graphene, silicene (A, B) and MoS$_2$ (C, D). We only include intravalley scattering in the present analysis. The same $\mathbf{q}$-mesh is used as for the scattering rate analysis. For the $\mathbf{k}$-space we evaluate the scattering rate in the $\mathbf{k}$-points that contribute, from a very dense $1500\times1500$ Monkhorst-Pack sampling of the first Brillouin zone, up to a given Fermi-level/carrier-density. As an example, the resulting $\mathbf{k}$-points are illustated by markers in Fig.~\ref{fig:RateGraphene} in the case of graphene. The resulting number of $\mathbf{k}$-points treated within the valley is 316, 378 and 514 for graphene, silicene and MoS$_2$, respectively.

The phonon scattering limited mobilities calculated here shows that graphene can have a mobility close to $10^6\,\mathrm{cm}^2/\mathrm{V}\,\mathrm{s}$ at $100\,\mathrm{K}$ and a carrier density of $3\times 10^{11}\mathrm{cm}^{-2}$, cf. Fig.~\ref{fig:mobility}.
At $300\,\mathrm{K}$, we obtain a mobility decreasing from $145,000\,\mathrm{cm}^2/\mathrm{V}\,\mathrm{s}$ at a carrier density of $1\times 10^{12}\mathrm{cm}^{-2}$
to $55,000\,\mathrm{cm}^2/\mathrm{V}\,\mathrm{s}$ at a carrier density of $3\times 10^{12}\mathrm{cm}^{-2}$.
In comparison, experiments have so far achieved room temperature values decreasing from roughly $90,000\,\mathrm{cm}^2/\mathrm{V}\,\mathrm{s}$ to $45,000\,\mathrm{cm}^2/\mathrm{V}\,\mathrm{s}$ at the same carrier densities.\cite{wang_one-dimensional_2013}
The same experiment also obtain mobilities up to $10^6\,\mathrm{cm}^2/\mathrm{V}\,\mathrm{s}$ at lower temperatures.
%At $40\,\mathrm{K}$ and a carrier density of $3\times 10^{12}\mathrm{cm}^{-2}$, the same experiment obtains a mobility of $1,000,000\,\mathrm{cm}^2/\mathrm{V}\,\mathrm{s}$ where we obtain $500,000\,\mathrm{cm}^2/\mathrm{V}\,\mathrm{s}$.
%up to $140,000\,\mathrm{cm}^2/\mathrm{V}\,\mathrm{s}$ at $300\,\mathrm{K}$ and a carrier density of $3\times 10^{12}\mathrm{cm}^{-2}$ and $10^6\,\mathrm{cm}^2/\mathrm{V}\,\mathrm{s}$ at $40\,\mathrm{K}$ and a carrier density of $4\times 10^{12}\mathrm{cm}^{-2}$.\cite{wang_one-dimensional_2013}
In addition, we find the mobility of silicene to be more than an order of magnitude lower than graphene but still very high.
We obtain a mobility of roughly $2100\,\mathrm{cm}^2/\mathrm{V}\,\mathrm{s}$ at $300\,\mathrm{K}$ and a carrier density of $3\times 10^{12}\mathrm{cm}^{-2}$.
Previous calculations have only considered ungated silicene and obtained values in the range of $10-1000\,\mathrm{cm}^2/\mathrm{V}\,\mathrm{s}$ at $300\,\mathrm{K}$.\cite{li_intrinsic_2013,wang_silicene_2013}
The only experiment on silicene presently published achieve a mobility of roughly $100\,\mathrm{cm}^2/\mathrm{V}\,\mathrm{s}$ at $300\,\mathrm{K}$.\cite{tao_silicene_2015}
For MoS$_2$ our calculated mobility decreases from a value of $1700\,\mathrm{cm}^2/\mathrm{V}\,\mathrm{s}$ at $100\,\mathrm{K}$ to approximately $400\,\mathrm{cm}^2/\mathrm{V}\,\mathrm{s}$ at $300\,\mathrm{K}$. These results are in good agreement with published first-principles simulations and experimental values.\cite{kaasbjerg_phonon-limited_2012,restrepo_first_2014,li_electrical_2015}
In comparison, experiments have achieved up to $200\,\mathrm{cm}^2/\mathrm{V}\,\mathrm{s}$ at $300\,\mathrm{K}$.\cite{radisavljevic_single-layer_2011}

% Meshed k-summary
We note that the electron-phonon coupling as a function of $\mathbf{q}$, as was shown for all materials for a single $\mathbf{k}$-point, was evaluated at all $\mathbf{k}$-points in the mobility calculations in Fig.~\ref{fig:mobility}. 
Another approach that is often applied is to evaluate the scattering rate along a single $\mathbf{k}$-line obtained from a $|g_{\mathbf{k}\mathbf{q}}^{\lambda n n'}|$ at a fixed $\mathbf{k}$-point. This can be transformed to a generic energy-dependence which is used to evaluate the mobility. This approach however, neglects part of the anisotropy in the bulk electron-phonon coupling, which was included in the present results.

\subsection{Intervalley scattering}
% Intervalley
\begin{figure}[!htbp]%[!t]%[!htbp]%
\centering
{\includegraphics[width=0.99\linewidth]{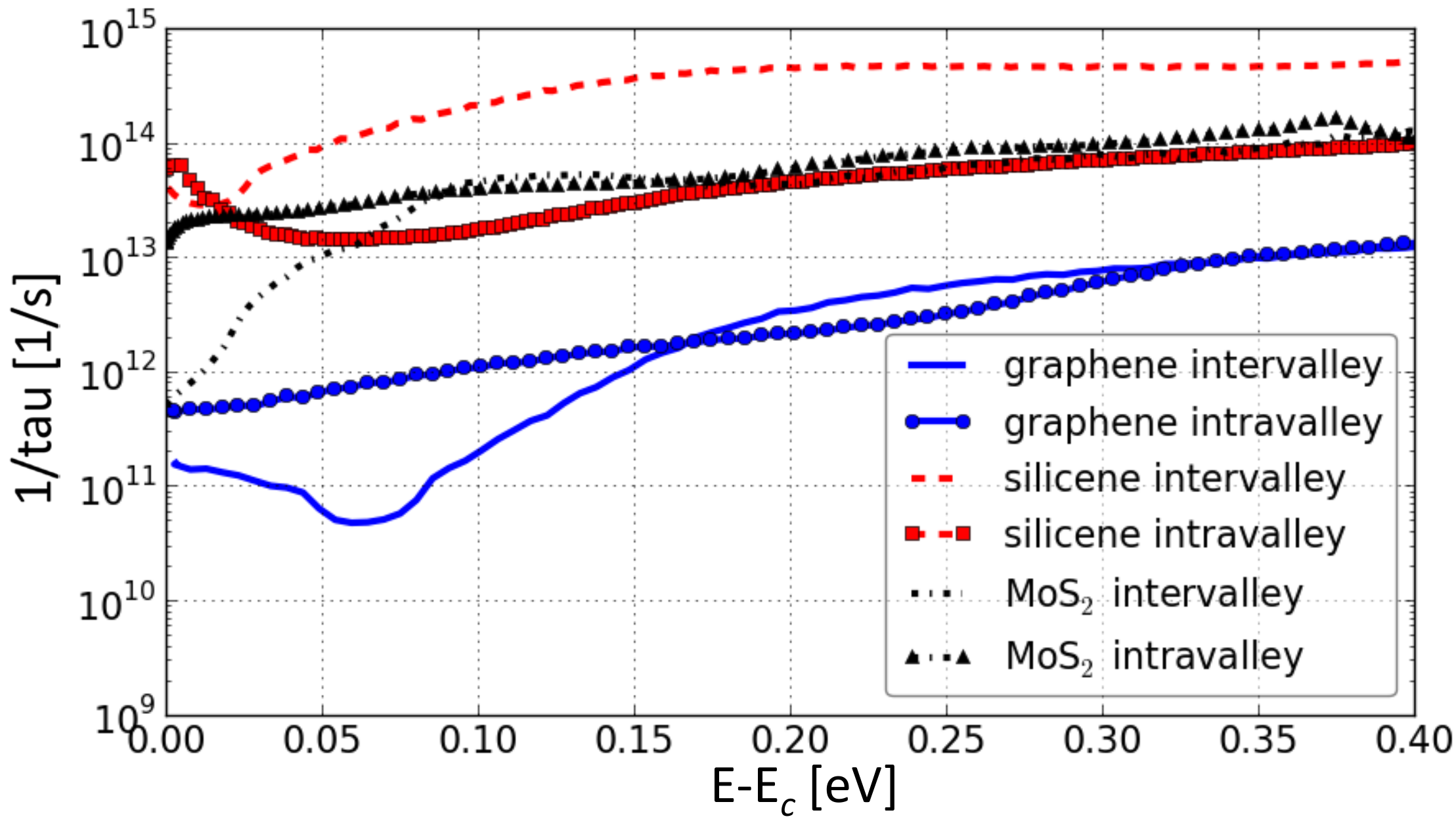}}%width=0.99\linewidth%width=2.5in
\caption{(Color online) Scattering rate as a function of energy/$\mathbf{k}$ for graphene, silicene and MoS$_2$ relative to the conduction band minimum $E_c$. In all cases the temperature is 300\,K. The Fermi-level and $\mathbf{k}$-route is the same as for the intravalley scattering rates presented in Fig.~\ref{fig:RateGraphene}B and Fig.~\ref{fig:RateSiliceneAndMoS2}.
%The Fermi-level is tuned to $\mu_F=100$\,meV for graphene and silicene and 500\,meV below the conduction band minimum for MoS$_2$. We used $\mathbf{k}$-points along a line from $K$ to $\Gamma$ up to 0.5eV for graphene and silicene and along a line from $K$ to $M$ for MoS$_2$.
}
\label{fig:IntervalleyScattering}
\end{figure}
We only included intravalley scattering in the previous figures. However, it is straightforward to evaluate the intervalley scattering rate separately by doing a calculation with the same $\mathbf{k}$-mesh and shifting the $\mathbf{q}$-mesh to $\mathbf{q}+\mathbf{K}$. We have done so for all three materials along a single $\mathbf{k}$-line to evaluate at what temperatures and carrier densities the intervalley scattering might start contributing.

In Fig.~\ref{fig:IntervalleyScattering} we compare the obtained intra- and intervalley scattering rates summed over all modes, except the intravalley ZA mode as discussed previously.
For graphene, we find that intervalley scattering starts contributing around $200-300\,$K and at a Fermi-level above $140$\,meV, corresponding to a carrier density of $n_0\approx 10^{12}\mathrm{cm}^{-2}$, cf. Fig.~\ref{fig:mobility}A. The reason is that intervalley scattering requires a phonon momentum connecting the two valleys $|\mathbf{q}|\approx|\mathbf{K}-\mathbf{K}'|\approx |\mathbf{K}|$ where the lowest energy phonon modes (ZA and ZO around $\mathbf{q}=\mathbf{K}$) only exist above $65\,$meV in the case of graphene, cf. Fig.~\ref{fig:Bandstructures}B.
A similar onset exist for silicene and MoS$_2$ where we at $300\,$K find from Fig.~\ref{fig:IntervalleyScattering} that intervalley scattering starts contributing at a Fermi-level of $30$\,meV and $100$\,meV above the conduction band minimum, respectively.
For MoS$_2$ this corresponds to $n_0\approx 10^{14}\mathrm{cm}^{-2}$ so that intervalley scattering does not contribute significantly, whereas for silicene intervalley scattering seems to make a significant contribution above $n_0\approx 0.5\times 10^{12}\mathrm{cm}^{-2}$.

Our definition of the onset for intervalley contribution is that we obtain an intervalley scattering rate of the same order of magnitude as for the intravalley scattering.
Using Matthiessen's rule $1/\mu=1/\mu^{intra}+1/\mu^{inter}$ we can estimate that the total mobility will decrease by approximately a factor of two at carrier densities above $n_0\approx 10^{12}\mathrm{cm}^{-2}$ if intervalley scattering was included for graphene. For MoS$_2$ the contribution will be much smaller at assessable carrier densities, whereas silicene can have as much as an order of magnitude lower mobility due to intervalley scattering for carrier densities above $n_0\approx 0.5\times 10^{12}\mathrm{cm}^{-2}$.

%In addition, the mobility 
%\newpage
\section{Conclusion\label{Sec:Conclusion}}
In summary, we have presented a Boltzmann Transport Equation (BTE) solver implemented in the Atomistix ToolKit (ATK) simulation tool.
The method allows for calculation of material properties, including the electron-phonon interaction, from first-principles.
We have applied the tool to calculate the phonon-limited mobility in n-type monolayer graphene, silicene and MoS$_2$.
Our results compares well to published theoretical results and experiments for MoS$_2$ and graphene and extends previous theoretical calculations for silicene.
The bulk electron-phonon coupling is highly anisotropic due to scattering suppression in different $q$-directions, as well as a nontrivial $q$-dependence, related to the combined symmetry of the electron and phonon states.
The simulations provide an upper bound for the electron mobilities of the selected 2D materials. %The simulations provide a reasonable
The \textit{ab initio} approach demonstrated in this paper can be directly applied to other materials in both 1D, 2D and 3D, larger nanostructures and may be straightforwardly extended to study cases with electron-impurity scattering.
In addition, we have illustrated how the reduction of the lattice symmetry, when going from graphene to silicene with no planar symmetry and further on to MoS$_2$ with no inversion symmetry, leads to additional scattering mechanisms and modes interacting with the charge carriers.
This makes a strong case for the need of a fully numerical solution of the BTE together with a first-principles method to evaluate the bulk electron-phonon interaction - especially, when going to systems with many modes, such as systems with low lattice symmetry, slab structures and nanostructured materials. 

\section*{Acknowledgment}
We thank K. Kaasbjerg (DTU) and  T. Frederiksen (San Sebastian) for useful discussions on the methods employed.
The authors acknowledges support from Innovation Fund Denmark, grant Nano-Scale Design Tools for the Semiconductor Industry (j.nr. 79-2013-1).
The Center for Nanostructured Graphene (CNG) is sponsored by the Danish Research Foundation, Project DNRF58.

\appendix
\section{Finite difference derivative of Hamiltonian} \label{HamiltonianDerivative-appendix}
The derivative of the Hamiltonian operator is calculated in a way similar to the dynamical matrix following the approach of Kaasbjerg et al. \cite{kaasbjerg_phonon-limited_2012}. We construct a supercell by repeating the primitive configuration a number of times in each periodic direction. Subsequently we displace the atoms in the central cell by $\pm \delta$ in all Cartesian directions (forward and backward) and calculate the derivative with respect to e.g. atom $i$ in the $x$-direction as
\begin{eqnarray} 
\frac{\partial \hat{H}}{\partial x_i} \approx \frac{\hat{H}(x_i+\delta) - \hat{H}(x_i-\delta)}{2\delta}\,, \label{eq:dHdR-finite-diff}
\end{eqnarray} 
where $\hat{H}(x_i+\delta)$ indicates the Hamiltonian operator obtained for the configuration where atom $i$ is displaced by $\delta$ in the positive $x-$direction. Note that Eq. \eqref{eq:dHdR-finite-diff} applies for the Hamiltonian operator, and not for the Hamiltonian matrix expressed in a basis of LCAO basis functions. In the latter case it is necessary to correct also for the displacement of the basis orbitals\cite{frederiksen_inelastic_2007}. 

The Hamiltonian operator has several terms:
\begin{eqnarray}
\hat{H} = \hat{T} + \hat{V}_{local} + \hat{V}_{NL}\,,
\end{eqnarray}
where $\hat{T}$ is the kinetic energy operator, $\hat{V}_{local}$ is the local potential including the exchange-correlation potential, the Hartree potential as well as the local pseudo-potential, while $\hat{V}_{NL}$ is the non-local Kleinman-Bylander pseudo-potential. Since the kinetic energy does not depend on the atomic coordinates the derivative of that is zero. The derivative of the local potential can be directly evaluated using Eq. \eqref{eq:dHdR-finite-diff}. The non-local potential is written as 
\begin{eqnarray}
\hat{V}_{NL} = \sum_i\sum_{\alpha\beta} |\chi_{\alpha}^i \rangle v_{\alpha \beta}^i \langle \chi_{\beta}^i|\,, \label{eq:V-NL}
\end{eqnarray}
where $|\chi_\alpha^i\rangle$ is a projector function centered on atom $i$ and $v_{\alpha \beta}^i$ are (fixed) projector coupling elements. The derivative is written as
\begin{eqnarray}
\frac{\partial \hat{V}_{NL}}{\partial x_j} &=& \frac{\partial}{\partial x_j} \sum_i\sum_{\alpha\beta} |\chi_{\alpha}^i \rangle v_{\alpha \beta} \langle \chi_{\beta}^i| \nonumber \\
&=&  \sum_{\alpha\beta} \left( \frac{\partial |\chi_{\alpha}^j \rangle }{\partial x_j} v_{\alpha \beta} \langle \chi_{\beta}^j| + 
 |\chi_{\alpha}^j \rangle v_{\alpha \beta} \frac{\partial \langle \chi_{\beta}^j|  }{\partial x_j} \right)\,.
\end{eqnarray}
The derivatives of the projector functions are evaluated numerically as in Eq. \eqref{eq:dHdR-finite-diff}.

Having calculated the derivative of the Hamiltonian operator we evaluate the derivative in the LCAO basis
\begin{eqnarray} 
\left(\frac{ \partial \mathbf{H}}{\partial x_i} \right)_{\mu\nu} = \langle \phi_\mu | \frac{\partial \hat{H}}{\partial x_i} | \phi_\nu \rangle \,. \label{eq:dHdR-matrix}
\end{eqnarray}

%\bibliography{BibTexs/BibOther2DMaterials,BibTexs/a-Graphene,BibTexs/a-BibDiverse,BibTexs/BibGrapheneElectron-phononcoupling,BibTexs/BibElectron-Phononcoupling,BibTexs/books,BibTexs/BibBoltzmannModels,BibTexs/BibSiNWs,BibTexs/BibBoltzmannTransportEquationAndMethods,BibTexs/BibOwnPapers,BibTexs/BibCID_Theory,BibTexs/a-BibDFTandSiestaMethods}
%

\end{thebibliography}

\end{document}